\documentclass[prl,twocolumn,10pt,longbibliography,amsmath,amssymb,aps,
]{revtex4-2}

\usepackage{amsmath}
\usepackage{amssymb}
\usepackage{amsfonts}
\usepackage{color}
\usepackage{graphicx}
\usepackage[hidelinks]{hyperref}
\usepackage[utf8]{inputenc}
\usepackage[english]{babel}
\usepackage{dsfont}
\usepackage{xspace}
\usepackage[normalem]{ulem}
\usepackage{tabularx}
\usepackage{xcolor}
\usepackage{csquotes}
\usepackage{mathtools}

\usepackage[caption=false]{subfig}

\newcolumntype{Y}{>{\centering\arraybackslash}X}
\newcolumntype{L}[1]{>{\raggedright\let\newline\\\arraybackslash\hspace{0pt}}m{#1}}
\newcolumntype{C}[1]{>{\centering\let\newline\\\arraybackslash\hspace{0pt}}m{#1}}
\newcolumntype{R}[1]{>{\raggedleft\let\newline\\\arraybackslash\hspace{0pt}}m{#1}}

\def\bea{\begin{eqnarray}}
\def\eea{\end{eqnarray}}

\def\ba{\begin{array}}
	\def\ea{\end{array}}

\def\Tr{\text{Tr}}
\def\sgn{\text{sgn}}

\def\avg#1{\left\langle#1\right\rangle}

\def\ket#1{\left|#1\right\rangle}

\def\abs#1{\left|#1\right|}
\def\kc#1{\left(#1\right)}
\def\kd#1{\left[#1\right]}
\def\ke#1{\left\{#1\right\}}

\def\sgn{{\rm sgn}}

\def\be{\begin{equation}}       \def\ee{\end{equation}}
\def\bea{\begin{eqnarray}}      \def\eea{\end{eqnarray}}
\def\ba{\begin{array}}
	\def\ea{\end{array}}
\def\bnum{\begin{enumerate} }
	\def\enum{\end{enumerate}}

\def\=>{\Rightarrow}
\def\>{\rightarrow}

\def\eye2{Fathbb{I}}

\def\tr{\mathrm{tr}}
\def\Tr{\mathrm{Tr}}

\newcommand{\eqmaintext}[1]{Eq.~(#1) of the main text}

\usepackage[compat=1.1.0]{tikz-feynman}
\newcommand{\twoptf}[1]{
	\feynmandiagram [inline=(a.base), layered layout, horizontal=a to b, small] {
		#1 ,
	};}

\begin{document}

\preprint{APS/123-QED}

\title{Entanglement in interacting Majorana chains\texorpdfstring{\\}{} and transitions of von Neumann algebras}

\author{Pablo Basteiro}
\author{Giuseppe Di Giulio} \email{giuseppe.giulio@uni-wuerzburg.de}
\author{Johanna Erdmenger}
\author{Zhuo-Yu Xian}
\affiliation{Institute for Theoretical Physics and Astrophysics and W\"urzburg-Dresden Excellence Cluster ct.qmat, Julius Maximilians University W\"urzburg, Am Hubland, 97074 Würzburg, Germany}

\date{\today}

\begin{abstract}
We consider Majorana lattices with two-site interactions consisting of a general function of the fermion bilinear. The models are exactly solvable in the limit of a large number of on-site fermions. The four-site chain exhibits a quantum phase transition controlled by the hopping parameters and manifests itself in a discontinuous entanglement entropy, obtained by constraining the one-sided modular Hamiltonian. Inspired by recent work within the AdS/CFT correspondence, we identify transitions between types of von Neumann operator algebras throughout the phase diagram. We find transitions of the form II$_1\leftrightarrow\,$\,III$\,\,\leftrightarrow\,\,$I$_\infty$ that reduce to II$_1\leftrightarrow\,\,$I$_\infty$ in the strongly interacting limit, where they connect non-factorized and factorized ground states.
Our results provide novel realizations of such transitions in a controlled many-body model.
\end{abstract}

\maketitle

\paragraph*{Introduction.---}

Entanglement in many-body systems and quantum field theories (QFTs) has recently been explored via a novel take on local operator algebras and axiomatic QFT \cite{Murray1936,Murray1937,vNeumann1940,Murray1943,Takesaki1979,Haag1996} (see~\cite{Witten:2018zxz,Sorce:2023fdx} for a review). A fruitful platform for these analyses is the Anti-de Sitter/Conformal Field Theory correspondence \cite{Maldacena1997,Witten,GubserKlebanovPolyakov} (also known as holography), which relates strongly coupled QFTs in $d$ dimensions with gravity theories on negatively curved spacetimes in $d+1$-dimensions. In this context, operator algebras, and especially \textit{von Neumann} algebras, have recently been leveraged to rigorously describe the entanglement structure of holographic systems \cite{Leutheusser:2021qhd,Leutheusser:2021frk,Witten:2021jzq,Witten:2021unn,Chandrasekaran:2022eqq,Chandrasekaran:2022cip,Banerjee:2023eew,Engelhardt:2023xer}. A direct consequence of these investigations is the algebra typification of the two  phases in the Hawking-Page transition \cite{Leutheusser:2021qhd,Leutheusser:2021frk}, characterized by factorized and non-factorized Hilbert spaces, respectively.\\
This state of the art motivates us to study transitions of operator algebras arising in interacting many-body quantum systems whose Hilbert space structure allows for a controlled analysis of entanglement. A useful object which helps in classifying types of algebras is the \textit{one-sided modular Hamiltonian} associated to a given subregion \cite{Haag1996}. For free fermionic systems this object is uniquely determined by the two-point correlation functions restricted to the subregion \cite{IngoPeschel_2003}. We extend these results to interacting fermionic systems by constraining the form of the one-sided modular Hamiltonian in the limit of a large number of on-site fermions. For a wide class of interacting Majorana lattices, we exploit this extension to identify the operator algebras underlying our models, together with their transitions between different regimes of the phase diagram. This paves the way for addressing the classification of algebras, and possible transitions thereof, in previously suggested models for discrete holography, such as $O(N)$-invariant aperiodic spin chains \cite{Basteiro2023}.\\
More precisely, in this Letter, we introduce a lattice model with $N$ Majorana fermions on each site, interacting via a general potential involving multi-body hoppings. In the large $N$ limit, all higher-point functions factorize. Thus, we solve the system exactly by obtaining the two-point correlation function for generic interaction potentials. We showcase the wide applicability of our techniques by considering instances of the model including both finite and infinite chains with nearest-neighbor hopping.\\
We report three main results: First, we derive the entropy of a two-site chain with generic interaction potential. This entropy is fully determined by the correlations between the two sites, which are dictated by the interaction potential via self-consistency. This result can be interpreted both as the thermal entropy of the chain at a finite temperature or as the entanglement entropy of a two-site subregion in a larger chain. Remarkably, we find that the entropy itself does not depend explicitly on the chosen potential. Second, for a four-site chain, we identify two phases of the system characterized by strong and weak correlations within any two-site subsystem of the chain relative to all other correlations in the system. In particular, we identify a regime where the correlation structure indicates the factorization of the ground state. For a vast class of potentials, we show that these phases are connected by a quantum phase transition. Moreover, we determine the entanglement entropy of the half-chain by imposing a constraint on the one-sided modular Hamiltonian. In correspondence with the phase transition, the entanglement entropy exhibits a discontinuity above a critical value of the interaction strength. Third, exploiting the exact solvability of our model, we identify transitions between the local operator algebras underlying the four-site chain. When the correlations within a given subregion are the most relevant in the system, we find a von Neumann algebra of type I$_{\infty}$, which in general encodes finite entanglement. In our model, entanglement vanishes, thus signaling factorization. In the opposite regime, the algebra is of type II$_{1}$, associated with infinite entanglement entropy, while still allowing for the definition of a trace functional. The intermediate domain is described by type III algebras, where the entanglement is infinite and a trace functional cannot be defined. Strikingly, in the strongly interacting limit and for an exponential potential in the fermion bilinear, we find that the transition between types II$_1$ and I$_{\infty}$ algebras occurs in correspondence with the identified quantum phase transition. This transition then connects a non-factorized ground state to a ground state factorized into a product state.

\paragraph*{Hamiltonian and Schwinger-Dyson equations.---}

We consider a lattice of $L$ sites with $N$ Majorana fermions $\psi^j_x$ at each site, with anti-commutation relations $\ke{ \psi^j_x, \psi^k_y}=\delta_{jk}\delta_{xy}$. The microscopic Hamiltonian is given by
\begin{equation}
    H= \frac{N}{2}\sum_{x, y=1}^L h_{xy} \kc{\frac2{\mathrm{i}N}\sum_{j=1}^N\psi_x^j\psi_y^j}\,,
    \label{eq:Hamiltonian}
\end{equation}
with a general interaction potential $h_{xy}$, which, without loss of generality, obeys $h_{xy}(\xi)=h_{yx}(-\xi)$ and $h_{xx}(\xi)=0$. The theory is invariant under a global $O(N)$ rotation $\psi_x^j\to \sum_k O^{jk}\psi_x^k$, where $O$ is an orthogonal matrix. It is a lattice counterpart of the Gross–Neveu model \cite{PhysRevD.10.3235} with a general interaction potential, and only the bubble diagrams contribute to two-point functions at leading order in $1/N$, see the Supplemental Material (SM) \footnote{See Supplemental Material [url] for more details.}. When $h_{xy}(\xi)\propto\xi^q$, the model is equivalent to the replicated Brownian Sachdev-Ye-Kitaev (SYK$_q$) model in disorder averaging \cite{SSS18rampSYK}. In this work, $h_{xy}(\xi)$ always includes a linear term in $\xi$.\\
We solve the model \eqref{eq:Hamiltonian} by introducing the effective action of two auxiliary bi-local fields, the Green's function $G_{xy}(\tau_1,\tau_2)=1/N\sum_j\psi^j_x(\tau_1)\psi^j_y(\tau_2)$ and the self-energy $\Sigma_{xy}(\tau_1,\tau_2)$, introduced as a Lagrange multiplier, in the spirit of \cite{SachdevYePRL1993,kitaev,Maldacena:2016hyu}. Here, $\tau$ denotes Euclidean time. We consider the canonical ensemble at temperature $1/\beta$ and write the thermal partition function $Z=\int \mathcal{D}\tilde{G}\mathcal{D}\tilde{\Sigma}e^{-S_E[\tilde{G},\tilde{\Sigma}]}$ with the effective action
\begin{eqnarray}
    -S_E/N~&=&\log \text{PF}(\partial_\tau \delta_{xy}-\Sigma_{xy})\nonumber\\
    ~&-&\frac12\sum_{x,y}\int_0^\beta d\tau_1 d\tau_2 G_{xy}(\tau_1-\tau_2)\Sigma_{xy}(\tau_1-\tau_2)\nonumber\\
    ~&-&\frac{1}{2} \sum_{x,y} \int_0^\beta d\tau h_{xy}(-2\mathrm{i}G_{xy}(0))\,,
    \label{eq:eff_action}
\end{eqnarray}
where $G_{xy}(\tau)=-G_{yx}(-\tau)$ and we have assumed time-translational invariance. From $Z$, we may derive the properties of the ground state (when $\beta\to\infty$), or the thermodynamics of the system at finite temperature. In the large $N$ limit, the saddle point approximation of the path integral leads to the \textit{Schwinger-Dyson} (SD) equations
\begin{align}
		&G_{xy}'(\tau_{12})-\sum_z\int d\tau_3 \Sigma_{xz}(\tau_{13})G_{zy}(\tau_{32})  =\delta_{xy}\delta(\tau_{12}),
		\label{eq:SD_eqs_G}\\
		&\Sigma_{xy}(\tau_{12}) =2\mathrm{i}\,h'_{xy}(-2\mathrm{i}G_{xy}(0))\delta(\tau_{12})\,,\label{eq:SD_eqs_Sigma}
\end{align}
with $\tau_{ij}\equiv\tau_i-\tau_j$, and $\Sigma_{xy}=-\Sigma_{yx}$ due to the conditions on $h_{xy}$ mentioned above. This model is exactly solvable in the large $N$ limit, in the sense that all higher-point functions factorize into products of two-point functions, i.e. we have large $N$ factorization.\\
We solve Eqs.~\eqref{eq:SD_eqs_G} and \eqref{eq:SD_eqs_Sigma} for general $\tau$ by leveraging the fact that the self-energy is solely determined by  $-2\mathrm{i}G_{xy}(0)$. We can thus solve for $G(0)$ by means of \textit{self-consistency} (SC) conditions. Using the form of \eqref{eq:SD_eqs_G} in Fourier space, SC imposes
\begin{equation}
    G_{xy}(0)=\frac1\beta\sum_{n} \left[(-\mathrm{i} \omega_n-\Sigma(\omega_n))^{-1}\right]_{xy}\,,
    \label{eq:self_consistency_condition}
\end{equation}
with $\omega_n=2\pi(n+1/2)/\beta$ and $\Sigma$ given by \eqref{eq:SD_eqs_Sigma}. The Green's function $G(\tau)$ is then obtained by inserting $G(0)$ back into \eqref{eq:SD_eqs_Sigma} and \eqref{eq:SD_eqs_G}.

\paragraph*{Two-site chain.---}

To provide an explicit application of our general techniques, we now focus on a system governed by a general Hamiltonian $H$ of the form \eqref{eq:Hamiltonian} with $L=2$ sites and at a finite temperature $1/\beta$. Without loss of generality, we can absorb $\beta$ into the general form of the Hamiltonian \eqref{eq:Hamiltonian}, i.e. set $\beta=1$. Thus, the density matrix reads
\begin{equation}
\label{eq:bchainDM}
    \rho=\frac{e^{-H}}{Z},\qquad H=Nh\kc{\frac{2}{\mathrm{i}N}\sum_j\psi_1^j\psi_2^j}\,,
\end{equation}
where $Z=\Tr(e^{-H})$ and $h(\xi)\equiv h_{12}(\xi)=h_{21}(-\xi)$. We can exactly solve the SD equation \eqref{eq:SD_eqs_G} to obtain the Green's function $G(\tau)$ \cite{Note1}. At $\tau=0$, the solution reads
\begin{equation}
	G_{xy}(0)=\frac{1}{2}
    \begin{pmatrix}
       1 & -\mathrm{i}\tanh{(h'(X))}
        \\
     \mathrm{i} \tanh{(h'(X))} & 1
\end{pmatrix}\,,
  \label{eq:solution_bchain_finteT_t0}
\end{equation}
with $X\equiv -2\mathrm{i}G_{12}(0)$. We can read off the SC equation
\begin{equation}
    X = -  \tanh{(h'(X))}\,,
    \label{eq:selfconsistency_bchain_finteT}
\end{equation}
which can alternatively be derived from \eqref{eq:self_consistency_condition}. In general, we have $-1< X<1$ regardless of the form of the potential. Two relevant regimes of \eqref{eq:selfconsistency_bchain_finteT} are when $\abs{X}\to0$ and $\abs{X}\to 1$, corresponding to weak and strong correlations between the two sites, respectively.\\
From the path integral in the large $N$ limit, we can study the thermodynamic properties of the system \cite{Note1}. In particular, we find the entropy density $S/N\equiv s$ to be 
\begin{align}
    s(X)=-\frac{1+X}{2}\log\frac{1+X}2-\frac{1-X}{2}\log\frac{1-X}{2}\,,
    \label{eq:entropy_from_DM}
\end{align}
where $X$ satisfies the SC equation \eqref{eq:selfconsistency_bchain_finteT}. The fact that the entropy density function \eqref{eq:entropy_from_DM} does not explicitly depend on the interaction potential signifies the first main result of this work. The free energy is given by $F=N( h(X)-s(X))+\mathcal{O}(1)$. Remarkably, we see that $S\to N\log 2+\mathcal{O}(1)$ in the strong correlation limit, and $S\to0$ in the weak correlation regime \cite{Note1}.

\paragraph*{Four-site chain:---}
 
We can think of the two-site system as being part of larger chains and take advantage of the results derived above to study the entanglement structure. As an example, we consider an open chain of length $L=4$ at zero temperature. Although our techniques are valid for any potential, we consider here a specific instance $h_{x,x+1}(\xi)=\mu_x \left(1-e^{J\xi}\right)/(2J)=h_{x+1,x}(-\xi)$, where $J>0$ is the interaction strength and $\mu_1=\mu_3\equiv\mu_a$,  $\mu_2\equiv\mu_b$ denote hopping parameters. All remaining entries of $h_{xy}$ are zero. For convenience, we introduce the hopping ratio $r\equiv \mu_a/\mu_b$. To access the ground state properties in the large $N$ limit, the hierarchy of parameters $N\gg\mu_b\beta,\mu_a\beta\gg1$ needs to be taken into account. We solve the SD equation \eqref{eq:SD_eqs_G} and obtain $G(\tau)$ \cite{Note1}, which at $\tau=0$ reads 
\begin{equation}
	G_{xy}(0)=\frac{1}{2}\left(
	\begin{array}{cccc}
		1 & \mathrm{i}\sin \theta & 0 &\mathrm{i}\cos \theta \\
		-\mathrm{i} \sin \theta & 1 &\mathrm{i}\cos \theta & 0 \\
		0 & -\mathrm{i} \cos \theta & 1 &\mathrm{i}\sin \theta \\
		-\mathrm{i} \cos \theta & 0 & -\mathrm{i} \sin \theta & 1 \\
	\end{array}
	\right)
 \label{eq:G_aba}
\end{equation}
where the parameter $\theta$ is determined by the SC constraint derived from \eqref{eq:self_consistency_condition},
\begin{equation}
   \frac{\tan{\theta}}{2}
   =\frac{G_{12}(0)}{2G_{23}(0)}
   = \frac{h_{12}'(-2\mathrm{i}G_{12}(0))}{h_{23}'(-2\mathrm{i}G_{23}(0)))}
    =r e^{J(\sin{\theta}-\cos{\theta})}\,.
    \label{eq:self_cons_eq_aba}
\end{equation}
This transcendental equation may be solved numerically and has a unique solution for $J<J_c$ and three solutions for $J>J_c$, where the critical value can be proven analytically to be $J_c=\sqrt{2}$ \cite{Note1}. This multi-valuedness of the SC equation indicates that the system exhibits a discontinuous behavior as a function of $r$, now to be seen as a control parameter.
\begin{figure}[t]
	\centering
	\includegraphics[width=\linewidth]{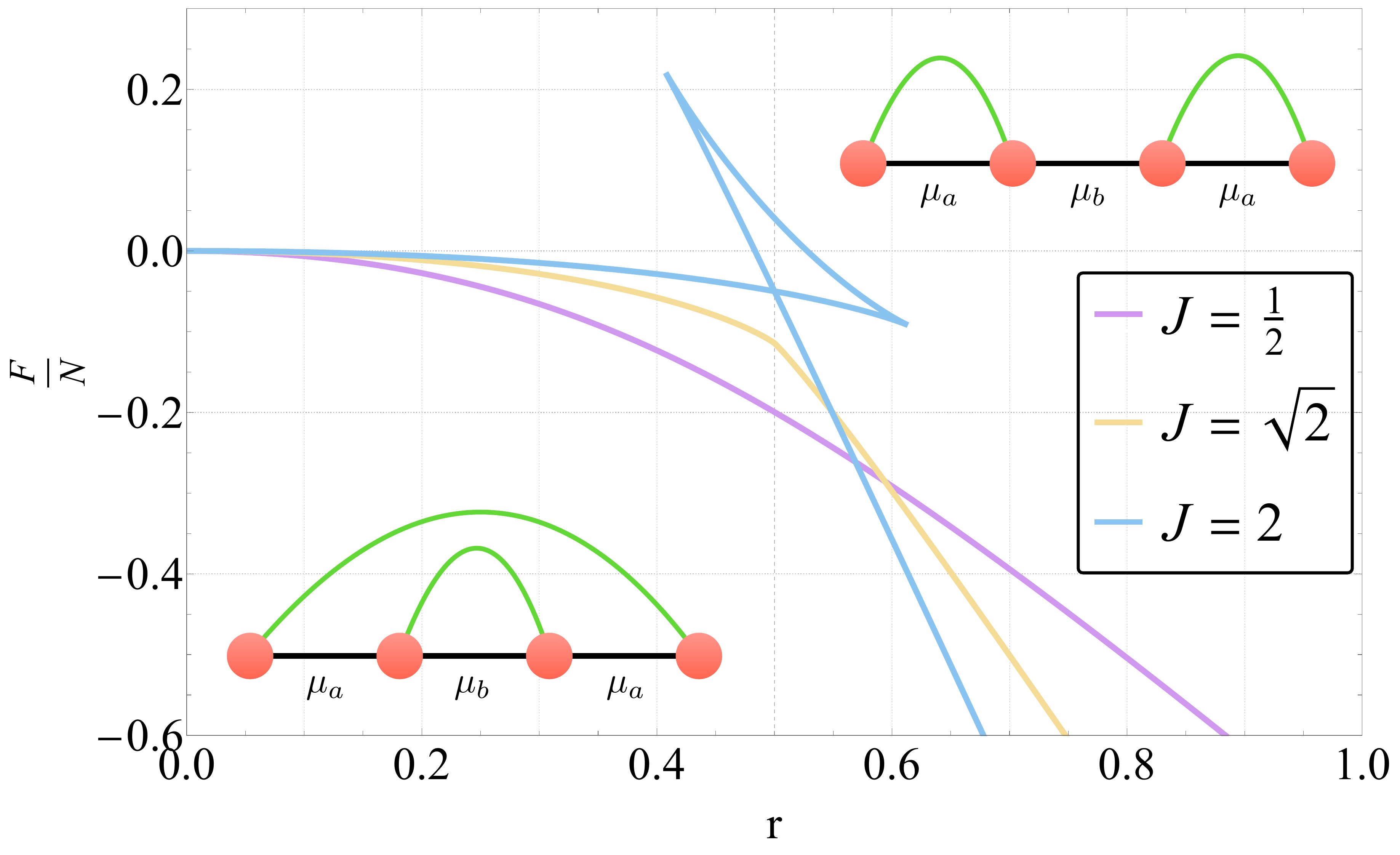}
	\caption{Free energy of the four-site chain with potential $h_{x,x+1}(\xi)=\mu_x \left(1-e^{J\xi}\right)/(2J)$ for different interaction strengths $J$. We observe a non-analyticity at $r\equiv\mu_a/\mu_b=1/2$ for $J$ above the critical value $J_c=\sqrt{2}$, signaling a phase transition. The two phases of the system are characterized by the correlation structure given by \eqref{eq:G_aba}, whose limiting cases for $r\to 0$ and $r\to \infty$ are shown in the two embedded diagrams.}
	\label{fig:free_energy_aba}
\end{figure}
We identify the thermodynamically dominant solutions by minimizing the free energy $F$ obtained from the effective action \eqref{eq:eff_action} \cite{Note1}. This free energy is shown in Fig.~\ref{fig:free_energy_aba} for different values of the interaction strength below, at, and above the critical point. We see that the free energy exhibits non-analyticity at $r=1/2$ for interaction strengths $J> J_c$. Thus, the system is characterized two phases for $J>J_c$ and it undergoes a first order quantum phase transition \cite{Sachdev_1999} across $r=1/2$, the existence of which constitutes the second main result of this work. At the critical point $J=J_c$, this transition is of second order. Let us stress that this phase transition is present for a large class of potentials other than the exponential \cite{Note1}. The two phases differ by the order parameter $\tan\theta$ \eqref{eq:self_cons_eq_aba}, which characterizes the correlation structure via \eqref{eq:G_aba}. Two limiting regimes of this structure when $r\to 0$ and $r\to\infty$ are shown in the insets of Fig.\ref{fig:free_energy_aba}.\\
We now turn our attention to the study of entanglement in the four-site model and consider a connected two-site subregion $A$, which we take to be, e.g., the sites $x=1,2$. The reduced density matrix $\rho_A$ of this system can be written as a thermal density matrix of a two-site chain of the form \eqref{eq:bchainDM}, with $H$ now to be thought of as the one-sided modular Hamiltonian. Its explicit form is not known in our case, so we take as an ansatz the general form given in \eqref{eq:Hamiltonian}. For this ansatz to describe a proper reduced density matrix, $\rho_A$ should reproduce the expectation values of local operators in the subregion. In particular, it must reproduce the correlations given by the Green's function \eqref{eq:G_aba} restricted to the subregion $A$. We must therefore impose a constraint for the one-sided modular Hamiltonian at large $N$, 
\begin{equation}
   G_{xy}(0) =\frac{1}{N}\sum_{j}\Tr\left(\rho_A \psi^j_x\psi^j_y\right)\,,\qquad
   x,y\in A\,,
 \label{eq:constraint_aba_chain_RDM}
\end{equation}
where $G_{xy}(0)$ is given in (\ref{eq:G_aba}). Eq.~\eqref{eq:constraint_aba_chain_RDM} uniquely determines the modular Hamiltonian only when $J=0$ \cite{IngoPeschel_2003}. Nevertheless, we can still use it to compute the entanglement entropy for $J\geq 0$. Indeed, we have shown that for a density matrix of the form given in \eqref{eq:bchainDM}, the entropy density of the two-site system can be computed for any form of the one-sided modular Hamiltonian ansatz and is given by \eqref{eq:entropy_from_DM}. Because of the constraint \eqref{eq:constraint_aba_chain_RDM}, $s(X)$ needs to be evaluated on the SC solution $X=-2\mathrm{i}G_{12}(0)=\sin\theta$ obtained from \eqref{eq:self_cons_eq_aba}, with $X\in[0,1)$ since $\theta\in[0,\pi/2)$. In this way, even without knowing the explicit form of $\rho_A$, we find that $s(X)=S_A/N$ is the entanglement entropy of subregion $A$.
\begin{figure}[t]
	\centering
	\includegraphics[width=\linewidth]{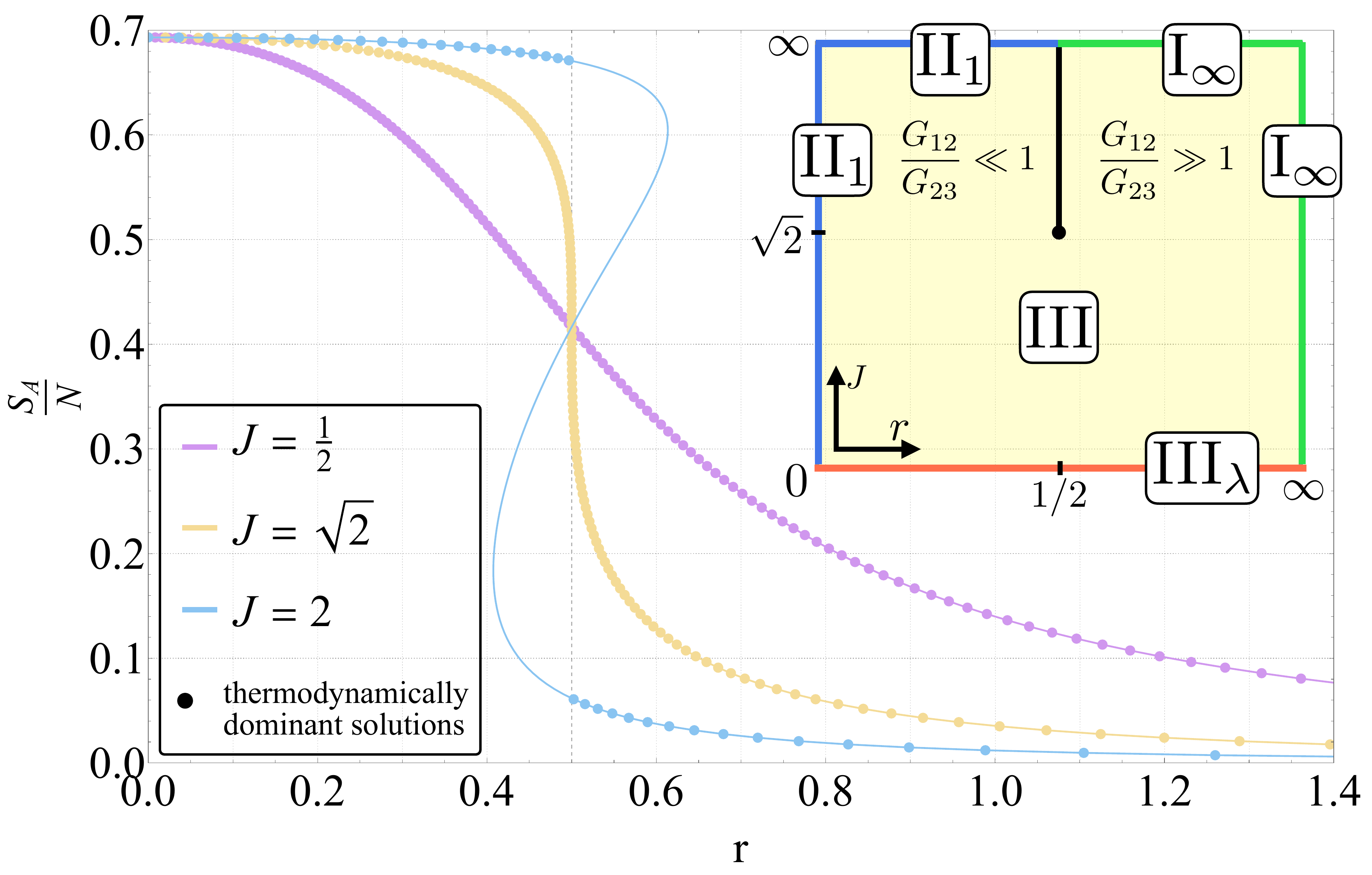}
	\caption{Entanglement entropy \eqref{eq:entropy_from_DM} of subregion $A$ on solutions to \eqref{eq:self_cons_eq_aba} as function of $r \equiv \mu_a/\mu_b$ for different coupling $J$. Solid lines denote all SC solutions, while a sample of physical solutions minimizing $F$ in Fig.~\ref{fig:free_energy_aba} is marked with dots. The phase transition is signaled by the discontinuity for $J>J_c$. Inset: von Neumann type of the algebra $\mathcal{A}_A$ in different regimes of the phase diagram. Each type is denoted by a different color, and the black dot (line) represents a phase transition of second (first) order.}
	\label{fig:entropy_aba}
\end{figure}
The resulting entanglement entropy density as a function of $r$ (recall $X$ depends on $r$ via \eqref{eq:self_cons_eq_aba}) is shown in Fig.~\ref{fig:entropy_aba} for different values of $J$. The phase transition reflects itself in the entanglement entropy evaluated on SC solutions as it becomes discontinuous for $J>J_c$.

\paragraph*{von Neumann algebras.---}

Motivated by recent results in holography \cite{Leutheusser:2021qhd,Leutheusser:2021frk,Witten:2021unn,Chandrasekaran:2022eqq,Chandrasekaran:2022cip,Banerjee:2023eew}, we study the classification of operator algebras associated to subsystems of our four-site model. Operator algebras can generally be classified into three types, denoted as type I, II and III \cite{Murray1936,Takesaki1979,Sorce:2023fdx}. Based on the standard trace Tr, a type I algebra encapsulates a finite entanglement entropy. Using Tr, entanglement entropy is infinite in both type II and type III algebras. To further distinguish the algebras, a key ingredient is the \textit{trace functional}, denoted by lower case \enquote{$\tr$} (to differentiate it from upper case Tr), which is defined to be a positive, linear and cyclic functional on the algebra \cite{Witten:2018zxz,Sorce:2023fdx}. In particular, type II algebras allow for the definition of such a trace functional, while type III algebras do not. For the technical construction of operators in the algebra, we closely parallel \cite{Witten:2018zxz}. In our system of consideration and in the $N\to\infty$ limit, operators in $\mathcal{A}_A$ consist of products of finitely many Majoranas located in subregion $A$. In particular, these will act trivially on countably infinitely many indices $j$ of the Majorana color space.\\
Based on the results for the ground state entanglement of our model, together with the operators in $\mathcal{A}_A$ defined above, we identify the operator algebras associated to subregion $A$ in different regimes of the correlation measure $X=-2\mathrm{i}G_{12}(0)$. This classification is shown in the phase diagram inset in Fig.~\ref{fig:entropy_aba}. When $X\to1$, the entropy $S_A\to0$ \cite{Note1}. This is consistent with our physical intuition since we expect the subsystems to completely factorize in this limit. Thus, we find that $\mathcal{A}_A$ is a type I$_{\infty}$ algebra when $X\to1$ (which implies $r\to\infty$ by \eqref{eq:self_cons_eq_aba}). The index in I$_{\infty}$ alludes to the infinite dimensionality of the local Hilbert space.\\
When $X<1$, the ground state is no longer factorized, and the entropy \eqref{eq:entropy_from_DM} is infinite, therefore ruling out $\mathcal{A}_A$ being of type I. To specify the type, we resort to the definition of a trace functional $\tr$ on $\mathcal{A}_A$. When the maximally entangled state $\ket{\Psi}$ is in the Hilbert space generated by the algebra $\mathcal{A}_A$, a well-defined trace functional is given by $\tr(\mathfrak{a})\equiv\langle\Psi|\mathfrak{a}|\Psi\rangle$, with $\mathfrak{a}\in\mathcal{A}_A$ \cite{Witten:2018zxz}. Importantly, when $X=0$, we find that the entanglement entropy in our ground state is infinite \textit{and} maximal up to subleading corrections in $1/N$ \cite{Note1}. This implies that our ground state can be mapped to $\ket{\Psi}$ by applying finitely many Majorana operators. Therefore, we conclude that the functional $\textrm{tr}$ defines a proper trace when $X=0$ ($r=0$ by \eqref{eq:self_cons_eq_aba}), thus unveiling that $\mathcal{A}_A$ is of type II$_1$ only at this point, cf.~Fig.~\ref{fig:entropy_aba}.\\
As for the regime $0<X<1$, corresponding to $0<r<\infty$ by virtue of \eqref{eq:self_cons_eq_aba}, we find that the entanglement entropy \eqref{eq:entropy_from_DM} is infinite but not maximal at leading order as $N\to\infty$. Therefore, our ground state cannot be mapped to the maximally entangled state $\ket{\Psi}$ by finitely many local operators, and therefore the algebra $\mathcal{A}_A$ does not admit the definition of a trace \cite{Witten:2018zxz}. This implies that $\mathcal{A}_A$ is of type III. Recall that the first and second equalities in \eqref{eq:self_cons_eq_aba} imposes SC for generic potentials, and therefore the analysis above is valid in the general interacting case.\\
In the free case $J=0$, where the entanglement Hamiltonian ought to be quadratic in the fermions, the classification of the algebras can be attained by studying the spectrum of the \textit{modular operator} $\Delta=\lim_{N\to\infty}\rho_A\otimes \rho_{\Bar{A}}^{-1}$ \cite{vNeumann1939,Powers1967,Araki1968,Takesaki1979}. Given this setup, we are able to compute the large-$N$ spectrum of $\Delta$ \cite{Note1}, finding $\textrm{Spec}(\Delta)=\{\lambda^n\}_{n\in\mathds{Z}}$. Here, the parameter $\lambda$ is related to the correlations within the subsystem as $\lambda=\frac{1-X}{1+X}$. When the modular operator has precisely this form, the associated operator algebras are said to be of I$_{\infty}$ when $\lambda=0$, type II$_1$ for $\lambda=1$, and type III$_{\lambda}$ for $\lambda\in(0,1)$. Such type III$_{\lambda}$ algebras are known to arise for free fermions on a lattice \cite{Pedersen2018}.\\
At finite $J$, these considerations lead to transitions between operator algebras of type II$_1\leftrightarrow\,$\,III$\,\,\leftrightarrow\,\,$I$_\infty$ in the phase diagram, cf.~Fig.~\ref{fig:entropy_aba}. Our setup's analytical tractability enables us to study the limit $J\to\infty$, where the solution to the SC equation \eqref{eq:self_cons_eq_aba} for the class of interaction potentials with exponential behavior is $X=\Theta(r-1/2)$, with $\Theta$ the Heaviside step function. In this limit, the entanglement entropy is $S_A=\Theta(1/2-r)N \log 2+\mathcal{O}(1)$ \cite{Note1}. This results in a direct transition of algebras II$_1\leftrightarrow\,$I$_\infty$ at $r=1/2$, which coincides with the phase transition undergone by the system. The transitions between different types of local operator algebras across the phase diagram provide the third result of our work.

\paragraph*{Closed periodic chains.---}

To showcase the generality of our methods, we now consider closed periodic chains. In particular, we focus on a closed chain consisting of $L$ sites with a Hamiltonian of the form \eqref{eq:Hamiltonian} with staggered interaction 
\begin{eqnarray}
\label{eq:staggeredinteraction}
        h_{xy}(\xi)&=&\delta_{x+1,y}(h_b(\xi)\delta_{\textrm{mod}_2 x,0}+h_a(\xi)\delta_{\textrm{mod}_2 x,1})\\&+&\delta_{x-1,y}(h_a(-\xi)\delta_{\textrm{mod}_2 x,0}+h_b(-\xi)\delta_{\textrm{mod}_2 x,1})\,,
        \nonumber
\end{eqnarray}
where $h_a$ and $h_b$ are generic functions and we have the periodic identification $L+x\sim x$. Notice that by defining \textit{cells} consisting of adjacent sites interacting by $h_b$, we can leverage translational invariance with respect to these cells to solve the model in momentum space \cite{Note1}. We find that the Green's function is determined by an implicit dependence on its own entries. In particular, $G$ is an implicit function of only the correlations within a given cell $G_{2x,2x+1}(0)$, and those connecting adjacent cells $G_{2x-1,2x}(0)$. This dependence manifests itself via the parametrization
\begin{align}
    \frac{1-v}{1+v}=\frac{h_a'(-2\mathrm{i}G_{2x-1,2x}(0))}{h_b'(-2\mathrm{i}G_{2x,2x+1}(0))}\,,
    \label{eq:SC_baba_ring}
\end{align}
where $v\in[-1,1]$. By the aforementioned translational invariance, the correlations entering \eqref{eq:SC_baba_ring} are independent of $x$. Explicit expressions for $G_{2x-1,2x}(0)$ and $G_{2x,2x+1}(0)$ at zero temperature and in the limit $L\to\infty$ can be found in terms of $v$ itself and read
\begin{align}
    &G_{2x-1,2x}(0)= \frac{\mathrm{i}}{2} g\kc{-v}\,,\quad 
    G_{2x,2x+1}(0) = \frac{\mathrm{i}}{2} g(v)\,,\label{eq:G_babaring}\\
    &g(v)=\frac{2\,\sgn(v)}{\pi}\kd{\frac{E(1-1/v^2)}{1+1/v}+\frac{K(1-1/v^2)}{1+v}}\,,
\end{align}
with $K(\xi),\,E(\xi)$ the complete elliptic integrals of the first and second kind, respectively. In the spirit of our methods, we can impose self-consistency inserting \eqref{eq:G_babaring} into \eqref{eq:SC_baba_ring} and solving for $v$. The model is thus completely solved once this value of $v$ has been found. Notice that $v\to-v$ exchanges $G_{2x-1,2x}(0)$ and $G_{2x,2x+1}(0)$ by virtue of \eqref{eq:G_babaring}, and this amounts to exchanging $h_a$ and $h_b$. Since $g(1)=1$, the limit $v\to 1$ and $v\to -1$ correspond to maximal correlations between nearest neighbor sites within and across cells, respectively. For periodic chains with $L=4$, we can parallel the previous discussions on the entanglement structure and, consequently, on the typification of the operator algebras across the phase diagram.

\paragraph*{Conclusions and future work.---}
 
We determine the phase structure and entanglement for a large class of Majorana models with $O(N)$ symmetry in the large $N$ limit. The von Neumann algebras underlying this entanglement structure are summarized in Fig.~\ref{fig:entropy_aba}. We provide a class of exactly solvable models that gives rise to non-trivial operator algebra transitions in a highly controllable way. This allows us to track the parameter regimes in which the correlations signal the factorization of the ground state into a product state, as shown in Fig.~\ref{fig:free_energy_aba}. While here we use correlations for characterizing factorization, a state-based approach consists of using entanglement orbits \cite{Banerjee:2023eew}. In spite of the different approaches used, we see a similar relation between factorization and the value of entanglement entropy. Analyzing these similarities is a promising line of future investigations.\\
Intriguingly, the algebra transition we find in the strong coupling limit $J\to\infty$ coincides with a phase transition that connects a factorized and a non-factorized state, similarly to the holographic Hawking-Page phase transition \cite{Leutheusser:2021qhd,Leutheusser:2021frk}. Differently from our case (i.e.~II$_1\leftrightarrow\,\,$I$_\infty$), this is a transition between algebras of type I$_{\infty}$ and III$_1$ as function of the temperature. Remarkably, we observe an analogous algebra transition although our model differs from previous works in the context of holography \cite{SachdevYePRL1993,kitaev,Sachdev:2015efa,Maldacena:2016hyu,Gu:2016oyy,Maldacena:2018lmt,Numasawa:2020sty}, which consider on-site random interactions leading to a non-zero entropy at zero temperature.\\
Further relations to holography can be obtained by including random disorder into our model by attaching an SYK model to each lattice site. This enlarges the phase diagram and we expect the competition between spatially inhomogeneous hoppings and the locally random disorder to change the renormalization group properties of critical points. Additionally, it is promising to investigate the mentioned connections with Brownian SYK \cite{SSS18rampSYK,StanfordScramblon23,milekhin2023revisiting,Jian:2021tli}.\\ 
Since our setup allows for general spatially disordered interactions, it is also relevant for further infinite disordered chains \cite{Vieira:2005PRB,Juh_sz_2007} where the hopping parameters are distributed according to a binary aperiodic sequence. These so-called aperiodic spin chains have recently been considered \cite{Jahn2020,Jahn:2021kti,Basteiro:2022zur,Basteiro2023} as a step towards establishing a holographic duality on discrete spaces.

\begin{acknowledgments}
\vspace{3em}
\noindent We are grateful to Souvik Banerjee, Moritz Dorband, Elliott Gesteau, Shao-Kai Jian, Changan Li, Ren\'e Meyer, Alexey Milekhin, and Sara Murciano for fruitful discussions. This work  was supported by Germany's Excellence Strategy through the W\"urzburg‐Dresden Cluster of Excellence on Complexity and Topology in Quantum Matter ‐ ct.qmat (EXC 2147, project‐id 390858490), and by the Deutsche Forschungsgemeinschaft (DFG) through the Collaborative Research Center \enquote{ToCoTronics}, Project-ID 258499086—SFB 1170, as well as a German-Israeli Project Cooperation (DIP) grant ”Holography and the Swampland”. ZYX also acknowledges support from the National Natural Science Foundation of China under Grant No.~12075298. We are grateful to the long term workshop YITP-T-23-01 held at the Yukawa Institute for Theoretical Physics, Kyoto University, where a part of this work was done.

\end{acknowledgments}

\newpage

\onecolumngrid
\renewcommand{\theequation}{S.\arabic{equation}}
\setcounter{equation}{0}

\section{Supplemental Material}

\subsection{Feynman diagrams}\label{apx:Feynman}

\begin{figure}[h!]
    \centering
    \includegraphics[width=0.65\textwidth]{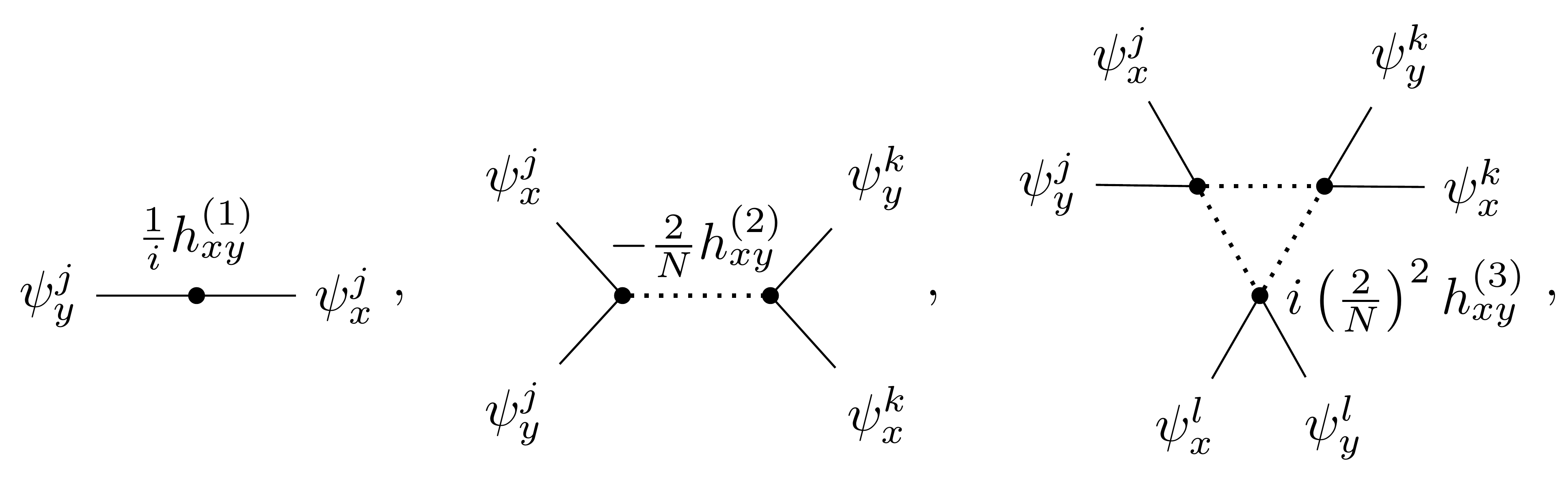}
    \caption{Interaction vertices for terms in \eqref{SMeq:hhhh} for orders $p=1,2,3$. Each dot indicates the non-local bilinear operator $\psi_x^j\psi_y^j$ and the dotted lines indicate the synchronicity of the interacting operators}
    \label{SMfig:feynman1}
\end{figure}

The Feynman diagrams are made of the bare propagator $G_0(\tau_1,\tau_2)=\frac12{\rm sgn}(\tau_1-\tau_2)$ and the simultaneous but non-local interactions given by the series expansion of the potential 
\begin{align}
    \frac N2\sum_{xy}h_{xy}\kc{\frac2{iN}\sum_j\psi_x^j\psi_y^j} 
    =\frac N2 \sum_{p=0} \sum_{xy} h_{xy}^{(p)} \kc{\frac2{iN}\sum_j\psi_x^j\psi_y^j}^p\,,
    \label{SMeq:hhhh}
\end{align}
where $h_{xy}^{(p)}$ are expansion coefficients. The $p=1,2,3$ order terms are represented by the interaction vertices shown in Fig.~\ref{SMfig:feynman1}.

\begin{figure}
    \centering
    \includegraphics[width=0.45\textwidth]{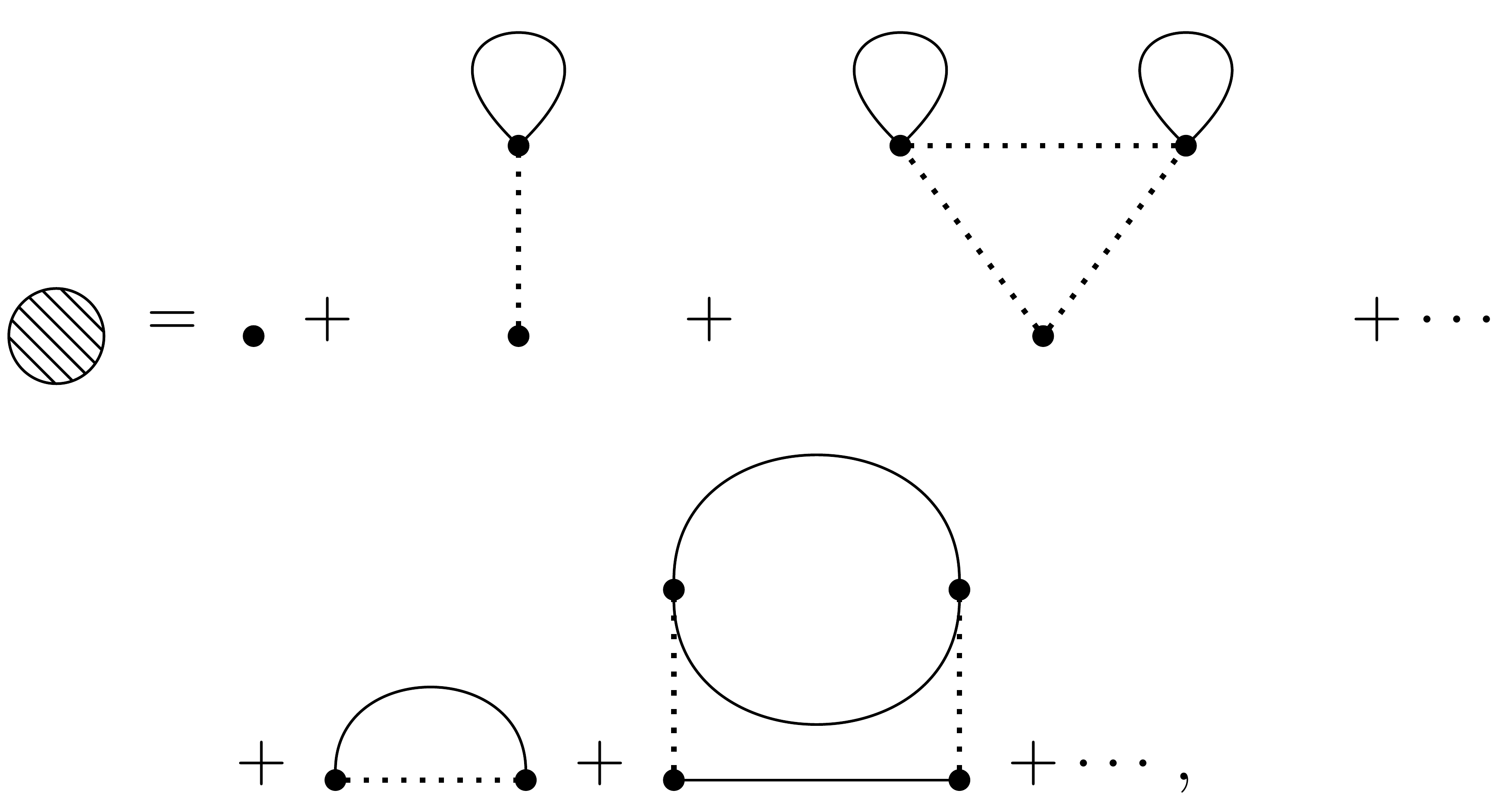}
    \caption{1PI diagram constructed from the diagrams in Fig.~\ref{SMfig:feynman1} through their combination and subsequent amputation of external legs.}
    \label{SMfig:feynman2}
\end{figure}

The Green's function $G_{xy}(\tau_1,\tau_2)=\avg{\frac1N\sum_j\psi_x^j\psi_y^j}$ in the presence of interactions is given by the iterative Feynman diagrams
\begin{align}
\twoptf{a -- b} \ = \ \twoptf{a -- [scalar] b} \ +\  \twoptf{a--[scalar] c [blob] -- b} \,,
\end{align}
where the dashed lines indicate the bare propagators $G_0$ and the blob indicates the one-particle irreducible (1PI) diagram. The self-energy $\Sigma(\tau_1,\tau_2)$ is given by the 1PI diagram shown in Fig.~\ref{SMfig:feynman2}.
From the summation over the color indices $j,k,l,\dots$, we can determine the order of the bubble diagrams in the $1/N$ expansion. The bubble diagrams in the first line of Fig.~\ref{SMfig:feynman2} are order $N^0$ and give rise to the Schwinger-Dyson (SD) equations (3) and (4) in the main text. Thus, all Green's functions reported in the main text receive contributions from these diagrams. The diagrams on the second line of Fig.~\ref{SMfig:feynman2} are order $N^{-1}$ and therefore are suppressed in the large $N$ limit. These provide sub-leading corrections to the results reported in the main text.

\subsection{Solutions of Schwinger-Dyson equations}\label{apx:solutions_SD}

\subsubsection{General case}

To solve the SD equations (3) and (4) in the main text, we work in Fourier space. For a system at finite temperature $\beta^{-1}$, we have
\begin{equation}
\label{eqSM:SDeq_Fourier}
   G_{xy}(\omega_n)=\left[\frac{1}{-\mathrm{i}\,\omega_n-\Sigma(\omega_n)}\right]_{xy}\,,
\end{equation}
where the Fourier transforms of $G(\tau_1,\tau_2)$ and $\Sigma(\tau_1,\tau_2)$ are defined by
\begin{equation}
   \label{eqSM:defFT_G}
G(\tau_1,\tau_2)=\sum_{n\in\mathds{Z}}e^{-\mathrm{i}\omega_n(\tau_1-\tau_2)}G(\omega_n)\equiv G(\tau_{12})\,,
\qquad\qquad
\Sigma(\tau_1,\tau_2)=\sum_{n\in\mathds{Z}} e^{-\mathrm{i}\omega_n(\tau_1-\tau_2)}\Sigma(\omega_n)\equiv\Sigma(\tau_{12})\,,
\end{equation}
and $\omega_n=2\pi(n+1/2)/\beta$ to account for the antiperiodicity of the fermionic variables. The dependence on the time difference $\tau_{12}\equiv \tau_1-\tau_2$ is due to the time-translational invariance. The entries of $G(\tau_{12})$ are obtained by taking the inverse Fourier transform of (\ref{eqSM:SDeq_Fourier}). In the limit of zero temperature, i.e. $\beta\to\infty$, the frequencies $\omega_n$ become a continuous variable $\omega$ and we replace $\sum_{n\in\mathds{Z}}\to\int_0^\beta d\omega$ in (\ref{eqSM:defFT_G}). In this work, we consider cases where the self-energy in \eqmaintext{4} depends on time through a delta function. This implies that $\Sigma(\omega)$ is independent of the frequency.

\vspace{-0.3cm}
\subsubsection{Two-site chain}

For the two-site chain discussed in the main text, the self-energy reads
\begin{equation}
\label{eqSM:2sites_SE0}
   \Sigma_{xy}(\tau)=
   2\mathrm{i}\delta(\tau)
   \begin{pmatrix}
        0 & -h'(X) \\
        h'(X) & 0  
    \end{pmatrix} \,,
\end{equation}
where we have chosen $h(\xi)\equiv h_{12}(\xi)=h_{21}(-\xi)$ in the Hamiltonian in \eqmaintext{1} and $X\equiv -2\mathrm{i}G_{12}(0)$. Inserting the Fourier transform of (\ref{eqSM:2sites_SE0}) into (\ref{eqSM:SDeq_Fourier}), we obtain
\begin{equation}
    \label{eqSM:SDeq_Fourier_2s}
   G_{xy}(\omega_n)=\frac{\mathrm{i}}{\omega_n^2+4[h'(X)]^2}
    \begin{pmatrix}
        \omega_n & 2h'(X) \\
       -2h'(X) &   \omega_n 
    \end{pmatrix}\,.
\end{equation}
Exploiting (\ref{eqSM:defFT_G}), we find the matrix of the two-point functions at finite temperature $\beta^{-1}=1$ for any value of $\tau>0$
\begin{equation}
\label{eqSM:2ptfunc-2sites}
    G_{xy}(\tau)= \frac{1}{2}
    \begin{pmatrix}
        \frac{e^{2h'(X)\tau}+e^{-2h'(X)(\tau-1)}}{1+e^{2h'(X)}} & \mathrm{i}\frac{\sinh{[h'(X)( 1-2\tau)]}}{\cosh{(h'(X))}}
        \\
     -\mathrm{i} \frac{\sinh{[h'(X)( 1-2\tau)]}}{\cosh{(h'(X))}} & \frac{e^{2h'(X)\tau}+e^{-2h'(X)(\tau-1)}}{1+e^{2h'(X)}}
    \end{pmatrix}\,,
\end{equation}
where the sums over $n$ in (\ref{eqSM:defFT_G}) for the various entries are evaluated by exploiting the standard trick of rewriting them as a contour integral of a function with simple poles along the imaginary axis. When $\tau=0$, (\ref{eqSM:2ptfunc-2sites}) becomes \eqmaintext{7}. Notice that $G(\tau)$ is not yet fully determined since it depends on the entry $G_{12}(0)$ itself. To conclude the derivation, we have to find the value of $G_{12}(0)$ as a function of the potential $h$ in the microscopic Hamiltonian. By comparing the off-diagonal components of both sides of (\ref{eqSM:2ptfunc-2sites}) at $\tau=0$, we obtain the self-consistency (SC) \eqmaintext{8}. The solution $G_{12}(0)$ obtained from the latter is to be plugged into (\ref{eqSM:2ptfunc-2sites}) to determine the Green's function for any time $\tau>0$.

\vspace{-0.3cm}
\subsubsection{Four-site chain}

Consider the four-site chain described by the Hamiltonian in \eqmaintext{1}, with a nearest-neighbor interaction potential given by $h_{x,x+1}(\xi)=h_{x+1,x}(-\xi)$ and all the other components of $h_{xy}(\xi)$ equal to zero. Notice that, although in the main text we have chosen a specific (exponential) form of the potential, our derivation holds for generic functions. In the following analysis, to preserve the symmetries of the example discussed in the main text, we further impose reflection symmetry, i.e. $h_{12}(\xi)=h_{34}(\xi)$.\\
The expression of the free energy is
\begin{equation}
    \Sigma_{xy}(\tau)=2\mathrm{i}\delta(\tau)\left(
\begin{array}{cccc}
 0  & -h'_{12}(-2\mathrm{i}G_{12}(0)) & 0 & 0 \\
 h'_{12}(-2\mathrm{i}G_{12}(0)) & 0  & -h'_{23}(-2\mathrm{i}G_{23}(0)) & 0 \\
 0 & h'_{23}(-2\mathrm{i}G_{23}(0)) & 0  & -h'_{12}(-2\mathrm{i}G_{12}(0)) \\
 0 & 0 & h'_{12}(-2\mathrm{i}G_{12}(0)) & 0  \\
\end{array}
\right)\,.
\label{SMeq:self_energy_four_sites}
\end{equation}
Inserting its Fourier transform in (\ref{eqSM:SDeq_Fourier}), we obtain
\begin{equation}
   G_{xy}(\omega_n)
    =
    \frac {\mathrm{i}}{b[\left(r^2+w_n^2\right)^2+w_n^2]}\left(
\begin{array}{cccc}
 w_n \left(r^2+w_n^2+1\right) & r \left(r^2+w_n^2\right) & r w_n & r^2 \\
 -r \left(r^2+w_n^2\right) & w_n \left(r^2+w_n^2\right) & w_n^2 & r w_n \\
 r w_n & -w_n^2 & w_n \left(r^2+w_n^2\right) & r \left(r^2+w_n^2\right) \\
 -r^2 & r w_n & -r \left(r^2+w_n^2\right) & w_n \left(r^2+w_n^2+1\right) \\
\end{array}
\right)
    \,,
    \label{SMeq:G_four_sites_full}
\end{equation}
where we define the shorthand notation $b\equiv 2 h_{23}'(-2\mathrm{i}G_{23}(0)))$, we notice that the dependence on the frequencies $\omega_n$ occurs through the new parameter $w_n\equiv\omega_n/b$ and we introduce the parametrization $r=\frac{h_{12}'(-2\mathrm{i}G_{12}(0))}{h_{23}'(-2\mathrm{i}G_{23}(0)))}=\frac12\tan\theta$ considered also in the main text. We are interested in the ground state correlations of the system, i.e. the limit $\beta\to\infty$. In this regime, the sums over the frequencies in (\ref{eqSM:defFT_G}) to determine the two-point functions in the time domain become integrals over $w\in[-\infty,\infty]$. We can perform these integrals, noticing that each entry of \eqref{SMeq:G_four_sites_full} has four poles given by 
\begin{equation}
   w_{\pm\pm}=\pm\frac{\mathrm{i}}{2} (\sec \theta \pm1) \,.
\end{equation}
If $\tau>0$, we choose the integration contour including the two poles with negative imaginary part. Using the residue theorem, the two-point functions are found to be
\begin{equation}
\label{eqSM:4site_Greenfunctions}
  G_{xy}(\tau)=\frac{e^{-\frac{1}{2} (s+1) b \tau } }{4s}
\left(
\begin{array}{cccc}
 (s+1) e^{b\tau }+s-1 & i \sqrt{s^2-1} \left(e^{b\tau }+1\right) & \sqrt{s^2-1} \left(e^{b\tau }-1\right) & i \left((s+1) e^{b\tau }-s+1\right) \\
 -i \sqrt{s^2-1} \left(e^{b\tau }+1\right) & (s-1) e^{b\tau }+s+1 & -i \left((s-1) e^{b\tau }-s-1\right) & \sqrt{s^2-1} \left(e^{b\tau }-1\right) \\
 \sqrt{s^2-1} \left(e^{b\tau }-1\right) & i \left((s-1) e^{b\tau }-s-1\right) & (s-1) e^{b\tau }+s+1 & i \sqrt{s^2-1} \left(e^{b\tau }+1\right) \\
 -i \left((s+1) e^{b\tau }-s+1\right) & \sqrt{s^2-1} \left(e^{b\tau }-1\right) & -i \sqrt{s^2-1} \left(e^{b\tau }+1\right) & (s+1) e^{b\tau }+s-1 \\
\end{array}
\right)  \,,
\end{equation}
where we introduce $s$ such that $\frac12\tan\theta=\frac12\sqrt{s^2-1}$, providing a more convenient parametrization for this result. As already seen for the two-site chain above, also in this case the Green's functions depend on their values at $\tau=0$ via the parameters $s$, $b$ and, in turn, $\theta$. By evaluating (\ref{eqSM:4site_Greenfunctions}) at $\tau=0$ and recalling the relation between $s$ and $\theta$, we obtain \eqmaintext{10}, which leads to the SC condition in Eq.\,(11). Inserting the solution of the SC condition back into (\ref{eqSM:4site_Greenfunctions}), we finally determine all the two-point functions in terms of the microscopic parameters of the Hamiltonian and for any value of $\tau>0$.

\subsection{Thermodynamics of two-site chain}\label{apx:b_chain_thermo}

\subsubsection{Free energy from path integral}\label{Apx:Free_energy_two_site}

We now derive the free energy $F$ of the two-site chain for generic interaction potentials. It is instructive to reinstate the physical temperature parameter $1/\beta$ for this purpose. We also write the interaction potential in the form $h'(\xi)\equiv\mu\,m(\xi)$ to extract the explicit dependence on the energy scale $\mu$, where $m(\xi)$ is dimensionless and encapsulates the functional dependence of the potential on the fermion bilinear. Using this notation, the Hamiltonian can be formally obtained by integration $h(\xi)=\mu\int d\xi'm(\xi')$. We further define the dimensionless quantity $\hat{\mu}\equiv \mu\beta$ for convenience. The free energy is obtained by evaluating the on-shell effective action on solutions to the SC equation given by \eqmaintext{8}. This is because from the path integral, we have
\begin{equation}
\label{SMeq:Partition_PI}   
Z=e^{-\beta F}=\int \mathcal{D}\tilde{G}\mathcal{D}\tilde{\Sigma}\,e^{- S_E[\tilde{G},\tilde{\Sigma},\hat{\mu}]}\,,
\end{equation}
where the tildes denote integration variables and $S_E$ is the Euclidean effective action given in \eqmaintext{2}, which we recall here for convenience
\begin{equation}
    \begin{split}
         -S_E/N~=\log \text{PF}(\partial_\tau \delta_{xy}-\Sigma_{xy})
    -\frac12\sum_{x,y}\int_0^\beta d\tau_1 d\tau_2 G_{xy}(\tau_1-\tau_2)\Sigma_{xy}(\tau_1-\tau_2)
    -\frac{1}{2} \sum_{x,y} \int_0^\beta d\tau h_{xy}(-2\mathrm{i}G_{xy}(0))\,.
    \label{SMeq:eff_action}
    \end{split}
\end{equation} 
Also, recall that $h(\xi)\equiv h_{12}(\xi)=h_{21}(-\xi)$, and the other components are zero, without loss of generality. Remembering that $X=-2\mathrm{i}G_{12}(0)$, we can write the third term of the effective action as
\begin{equation}
    -\frac{1}{2} \sum_{x,y} \int_0^\beta d\tau h_{xy}(-2\mathrm{i}G_{xy}(0))=-\frac{\beta}{2}\left(h_{12}(X)+h_{21}(-X)\right)=-\beta h(X)=-\hat{\mu}\int^X d\tilde{X}m(\Tilde{X})\equiv S_m[G](\hat{\mu})\,.
\end{equation}
Therefore, the effective action can be written as
\begin{equation}
   -S_E \equiv S_{\textrm{Pf}}[\Sigma]+S_G[G,\Sigma]+S_m[G](\hat{\mu})\,.
\end{equation}
In the large $N$ limit, we can approximate the path integral in \eqref{SMeq:Partition_PI} by its saddle-point value, which then allows us to write
\begin{equation}
   \log{Z}=-\beta F\approx - S_E[G,\Sigma,\hat{\mu}]\,,
\end{equation}
where now the action is to be evaluated on-shell, i.e. on the solutions to the SD equations (3) and (4) of the main text. Subsequently, $S_E$ is to be computed on solutions to the SC condition given by \eqmaintext{8}, which can be written as 
\begin{equation}
    \hat{\mu} = - \frac{\tanh^{-1}{(X)}}{m(X)}\,.
    \label{SMeq:two_site_SC}
\end{equation}
As discussed in \cite{Maldacena:2016hyu}, the direct evaluation of the effective action is non-trivial due to the presence of the Pfaffian determinant. To avoid this, we consider the derivative of the action with respect to $\hat{\mu}$,
\begin{equation}
    \begin{split}
        \hat{\mu}\frac{\partial}{\partial \hat{\mu}}\left(-S_E\right)
        &=\hat{\mu} \frac{\partial \Sigma}{\partial \hat{\mu}}\left(\frac{\delta S_{\textrm{Pf}}}{\delta \Sigma}+\frac{\delta S_{G}}{\delta \Sigma}\right)+\hat{\mu} \frac{\partial G}{\partial \hat{\mu}}\left(\frac{\delta S_{G}}{\delta G}+\frac{\delta S_{m}}{\delta G}\right)+\hat{\mu}\left(\frac{\partial S_m}{\partial \hat{\mu}}\right)\,.
    \end{split}
    \label{SMeq:logder_S}
\end{equation}
On-shell, the first two terms on the right hand side vanish and the only remaining contribution is
\begin{equation}
    \label{SMeq:logder_S_onshell}
    \frac{\partial}{\partial \hat{\mu}}\left(-S_E/N\right)_{\big|_{\textrm{\tiny on-shell}}}=\frac{\partial}{\partial\hat{\mu}}\left(S_m[G](\hat{\mu})/N\right)_{\big|_{\textrm{\tiny on-shell}}}=- \int^{X}d\tilde{X}m(\tilde{X})=-\frac{h(X)}{\mu}\,,
\end{equation}
where the on-shell condition imposes $X=-2\mathrm{i}G_{12}(0)$. In order to obtain back the free energy, we need to integrate the expression above with respect to $\hat{\mu}$, keeping in mind that, on-shell, the argument of the potential function $h(X)\equiv h(X(\hat{\mu}))$ is now implicitly dependent on the integration constant, with dependence given through \eqref{SMeq:two_site_SC}. However, we can circumvent this by changing the integration variable using $\frac{d\hat{\mu}}{\mu}=d\beta=\frac{\partial \beta(X)}{\partial X}dX$. From the SC equation \eqref{SMeq:two_site_SC}, and recalling that $h'(X)=\mu\,m(X)$, we see that $\frac{\partial \beta(X)}{\partial X}=\frac{\partial}{\partial X}\left(-\frac{\tanh^{-1}(X)}{h'(X)}\right)$. Therefore, we find the free energy to be given by
\begin{equation}
    \frac{\beta F}{N}=\int_{X_0}^{X} d\tilde{X} \left(\frac{\partial \beta}{\partial X}\right)_{|_{X=\tilde{X}}} h(\tilde{X})=-\frac{h(X) \tanh ^{-1}(X)}{h'(X)}+\frac{1}{2} \log \left(1-X^2\right)+X \tanh ^{-1}(X)\,,
    \label{eq:General_Onshell_Action_FiniteT}
\end{equation}
to be evaluated on those values of $X$ which are solutions to \eqref{SMeq:two_site_SC}. Notice that the expression of $F$ is correct up to a constant which is independent of $X$ and $\hat{\mu}$, which originate from those terms of \eqref{SMeq:logder_S} which give zero contribution on-shell. Additionally, the internal energy of the system is given by (cf.~\eqref{SMeq:logder_S_onshell})
\begin{equation}
    E=-\partial_{\beta}\log{Z}=\mu\partial_{\hat{\mu}}S_E\,\,\Longrightarrow\,\,\frac{\beta E}{N}=\frac{\beta h(X)}{N}\,.
\end{equation}
Computing the entropy $S$ at leading order in $1/N$, we find
\begin{equation}
    \begin{split}
         \frac{S}{N}=\frac{\beta E}{N}-\frac{\beta F}{N}&=-X\tanh^{-1}{X}-\frac{1}{2}\log(1-X^2)+\log 2+h(X) \left(\frac{\tanh ^{-1}(X)}{h'(X)}+\beta \right)\,,\\
    \end{split}
\end{equation}
where the $\log 2$ comes from physical considerations on the maximal and minimal values that a thermodynamic entropy ought to have, fixing in this way the additive constant of $F$. Remarkably, the expression in brackets in the last term is precisely the SC equation \eqref{SMeq:two_site_SC} and it therefore vanishes. We can then re-write the remaining contributions to the entropy density in the form
\begin{equation}
    \frac{S}{N}= -\frac{1-X}{2}  \log \left(\frac{1-X}{2}\right)-\frac{1+X}{2} \log \left(\frac{1+X}{2}\right)\,,
    \label{SMeq:entropy_function}
\end{equation}
which is the exactly same entropy density given in \eqmaintext{10}.

\subsubsection{Subleading corrections to the entropy}

In this section we calculate the subleading $1/N$ corrections to the entropy of the two-site system with density matrix
\begin{equation}
\label{SMeq:bchainDM}
    \rho=\frac{e^{-Nh(\chi)}}{Z}\,,\qquad \chi=\frac{2}{\mathrm{i}N}\sum_{j=1}^N \psi_1^j\psi_2^j\,.
\end{equation}
Notice that this density matrix is the one considered in \eqmaintext{6}. Recall that the function $h(\xi)$ satisfies the SC constraint
\begin{equation}
    X = -  \tanh{(h'(X))}\,,
    \label{SMeq:selfconsistency_bchain_finteT}
\end{equation}
with $X=-2\mathrm{i}G_{12}(0)$ given by the  Green's function of the system in the state (\ref{SMeq:bchainDM}). Notice that operator $\chi$ in (\ref{SMeq:bchainDM}) is related to the occupation number $n$ by $\chi=2n/N-1$, such that the Hamiltonian is diagonal on the occupation basis $ \ke{\ket{n_1,n_2,\cdots,n_N} | n_j\in \ke{0,1}, \forall j } $ with eigenvalue $Nh(2n/N-1)$, where $n=\sum_j n_j$. Then the partition function of the system is
\begin{align}
    Z=\sum_{n=0}^N 
    \begin{pmatrix}
        N \\ n
    \end{pmatrix}
    e^{-Nh(2n/N-1)}
    \to
    \int_{-1}^1 d\chi\, \sqrt{\frac{N}{2\pi  (1-\chi^2)}}\,e^{-Nf(\chi)}\,,
    \label{eq:part_funct_from_DM}
\end{align}
where we have fixed $0<n/N=(\chi+1)/2<1$ and sent $N\to\infty$ at the last step. Further, we have introduced the function
\begin{align}
\label{eqSM:hands_def}
     f(\chi)=h(\chi)-s(\chi)\,,\quad s(\chi)=-\frac{1+\chi}{2}\log\frac{1+\chi}2-\frac{1-\chi}{2}\log\frac{1-\chi}{2}\,.
\end{align}
For large $N$, the integral can be approximated by its saddle point value, which imposes $\chi=X$. This is consistent with the fact that $-1<X<1$ as seen from \eqref{SMeq:selfconsistency_bchain_finteT}.\\
To calculate the entropy in the saddle point approximation, we first define a functional of an arbitrary smooth function $Y(\chi)$ as
\begin{align}
    \avg{Y(\chi)}=\frac1{Z_0}\sqrt{\frac{N}{2\pi}}\int_{-\infty}^\infty d\chi\, Y(\chi)\,e^{-Nf(X)-\frac N2 f''(x)(\chi-X)^2}\,,\\ 
    Z_0=\sqrt{\frac{N}{2\pi}}\int_{-\infty}^\infty d\chi\,e^{-Nf(X)-\frac N2 f''(x)(\chi-X)^2}=\frac{e^{-Nf(X)}}{\sqrt{f''(X)}}\,.
\end{align}
Due to the Gaussian form of the integral, we have
\begin{align}\label{correlator}
    \avg{1}=1,\quad \avg{\chi-X}=0,\quad \avg{(\chi-X)^2}=\frac1{Nf''(X)}\,.
\end{align}
The functional can therefore be calculated via a series expansion $\avg{Y(\chi)}=\sum_n \frac1{n!}Y^{(n)}(X)\avg{(\chi-X)^n}$ and can be used to compute the thermodynamic quantities of the system. In the saddle point approximation, the partition function $Z$, the energy $E$, and the entropy $S$ are written as
\begin{align}
    Z=Z_0\avg{\frac1{\sqrt{1-\chi^2}}},\quad E=\frac{\avg{h(\chi)/\sqrt{1-\chi^2}}}{\avg{1/\sqrt{1-\chi^2}}},\quad S=E+\log Z\,.
    \label{SMeq:averaged_quantities}
\end{align}
The leading $1/N$ correction can be extracted from the $(\chi-X)^2$ order of the series expansions of the averaged quantities in \eqref{SMeq:averaged_quantities}. For the entropy, this correction is given by
\begin{align}
\label{eqSM:SandCorr}
    S= N s(X)+\frac{1}{2} \left(1+\log \frac{f''(X)}{1-X^2}+\frac{2 X h'(X)-1}{f''(X)(1-X^2)}\right)\,,
\end{align}
and is therefore of $\mathcal{O}((1/N)^0)$ for $\abs{X}<1$. Notice that, when $X\to 1$,  the subleading corrections to $S$ diverge and the computation above is not valid. The breakdown of the calculation can be physically interpreted as follows. The result (\ref{eqSM:SandCorr}) is obtained by assuming the system is characterized by the density matrix (\ref{SMeq:bchainDM}) and the Hamiltonian $h$. The form (\ref{eqSM:SandCorr}) is general as long as the system is in a mixed state and stops being valid only if the state becomes pure. Thus, the divergence of the corrections to $S$ in (\ref{eqSM:SandCorr}) suggests that when $X\to 1$, the system is in a pure state. Crucially, if the system is in a pure state, $S=0$ to all orders in the $1/N$ expansion. This is consistent with the fact that $\lim_{X\to 1}s(X)=0$, as discussed in the main text and noticed from (\ref{eqSM:hands_def}).\\
As explained in the main text, the corrections (\ref{eqSM:SandCorr}) and the fact that $S=0$ when $X\to 1$ can be exploited both as results for the thermal entropy of a two-site chain or for the entanglement entropy of a two-site block within a larger chain. The case of a two-site subsystem in a four-site chain was considered in the main text (see also the following sections of this SM). In that specific instance, we can provide further arguments supporting the fact that $S_A\to0 $ when $X\to 1$ (the change of notation for $S$ is due to the fact that in this example we always refer to an entanglement entropy). Indeed, we first recall from Eqs.~(10) and (11) of the main text that $X\to 1$ corresponds to $r=\mu_a/\mu_b\to \infty$. By setting $\mu_b=0$ in the Hamiltonian \eqmaintext{1}, and subsequently taking the large $N$ limit, we still obtain the SD equations (3) and (4) of the main text. In other words, the limits $\mu_a/\mu_b\to \infty$ and $N\to \infty$ commute. If $\mu_b=0$ in \eqmaintext{1}, it follows straightforwardly that $S_A=0$, given that the two parties of the bipartition factorize. Given the commutativity of the limits, we expect that this property also survives if we first take $N\to\infty$ and then $\mu_a/\mu_b\to \infty$, i.e. $X\to 1$. The above arguments also match the analysis from the Feynman diagrams. When considering $\mu_b\to0$, the self-energy $\Sigma_{xy}$ with $x=1,2$ and $y=3,4$, including the whole $1/N$ series expansion in Fig.~\ref{SMfig:feynman2}, vanish. It means that the correlations $G_{xy}$ with $x=1,2$ and $y=3,4$, including their $1/N$ corrections, also vanish in the limit $\mu_b\to0$. Therefore, we find that $S_A\to 0$ when $X\to 1$ at all orders in $1/N$.

Summarizing, in this section we showed that the subleading corrections to the entropy of a two-site density matrix in the large $N$ expansion are of order $(1/N)^0$ when $|X|<1$, while $S=0$ at all order in $1/N$ if $X\to 1$.
When representing the entanglement entropy of half of a four-site chain with Hamiltonian in \eqmaintext{1}, these results are crucial ingredients in the classification of the operator algebras in the various regimes of the phase diagram of the model.

\subsection{Ground state properties of four-site chain}

\subsubsection{Self-consistency equation and critical point}

\begin{figure}
    \centering
    \includegraphics[width=0.85\textwidth]{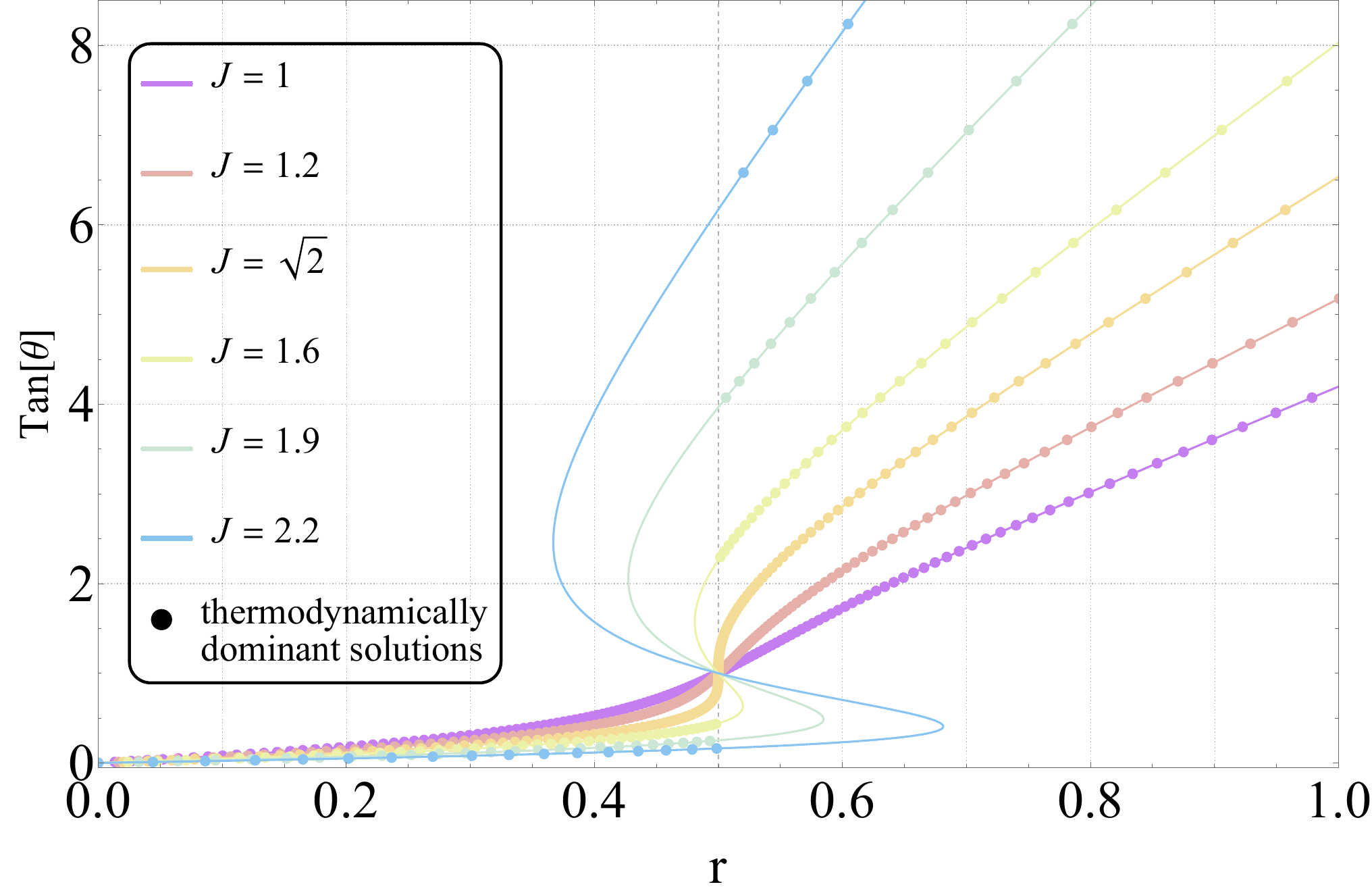}
    \caption{Solutions to the self-consistency equation \eqref{eq:Self_Cons_Eq} for an exponential potential with different interaction strengths, as function of the hopping ratio $r \equiv \mu_a/\mu_b$. Above the critical value $J=J_c=\sqrt{2}$, the equation starts having three possible solutions. Solid lines denote all SC solutions, while the dots represent a subset of all the thermodynamically dominant solutions, which are the ones minimizing the free energy \eqref{SMeq:free_energy_aba} (derived in the next section) and shown in Fig.~1 of the main text. For $J>J_c$ the SC solutions which are thermodynamically dominant exhibit a discontinuous jump at $r=1/2$, indicating that the systems undergoes a phase transition. The two phases are characterized by their correlation structure, as shown schematically in the insets of Fig.~1 of the main text. }
    \label{SMfig:tantheta_plot}
\end{figure}
This section provides a detailed analysis of the SC equation for the four-site chain given in \eqmaintext{11} with generic interaction potential obeying $h_{x,x+1}(\xi)=h_{x+1,x}(-\xi)$. We also define the short-hand notation $h_{x,x+1}(\xi)\equiv h_x(\xi)$. As in the previous section, it is instructive to extract the hopping parameters from the potential and define $h'_{x,x+1}(\xi)=\mu_x m(\xi)$, where $\mu_1=\mu_3=\mu_a$, and $\mu_2=\mu_b$ denote the hopping parameters on the four-site chain. We define the hopping ratio $\frac{\mu_a}{\mu_b}\equiv r$. Keep in mind that the function $m(\xi)$ implicitly contains any dependence on the interaction strength $J$. For concreteness, the reader can follow the analysis of this section with the exponential potential $m(\xi)=-e^{J \xi}/2$ in mind. The self-consistency equation for this potential is given by
\begin{equation}
    r\,e^{J(\sin{\theta}-\cos{\theta})}=\frac{1}{2}\tan{\theta}\,.
    \label{eq:Self_Cons_Eq}
\end{equation}
We plot solutions to this equation in Fig.~\ref{SMfig:tantheta_plot}, where we observe that all curves pass through the point $r=1/2$ and $\tan\theta=1$. This is a consequence of $\theta=\pi/4$ being a fixed point of the equation \eqref{eq:Self_Cons_Eq}, in the sense that $\sin(\pi/4)=\cos(\pi/4)$ and therefore the self consistency equation is automatically solved, independently of the value of $J$. We also identify a value $J\equiv J_c=\sqrt{2}$ below which there are unique solutions and above which the SC has three possible solutions. The fixed point of the SC equation mentioned previously allows us to find the critical value $J_c$ where the multi-valuedness of the solutions to the SC equation sets in for general potentials, as we show below.\\
In the general case, we can consider a generic potential $h'_x(\xi)$ with $x=a,b$ and $\xi$ either $\sin\theta$ or $\cos\theta$, as per \eqmaintext{11}. The self consistency equation \eqref{eq:Self_Cons_Eq} specifies the dependence of the hopping ratio $r$ as a function of $\theta$ as
\begin{equation}
    r(\theta)=\frac{1}{2}\tan\theta \frac{m(\cos\theta)}{m(\sin{\theta})}\,.
    \label{eq:rho_of_theta}
\end{equation}
Notice that now only the function $m$ appears, since the hopping ratio has been extracted explicitly into $r$. In order to identify the critical point, we need to investigate the extrema and inflection points of the function $r(\theta)$ defined in \eqref{eq:rho_of_theta}. We do this by setting its first and second derivative equal to zero. For general functions $m$, this would yield two equations with three unknowns, namely $m(\xi(\theta^*)),\,m'(\xi(\theta^*))$ and $m''(\xi(\theta^*))$, where $\theta^*$ is the point under consideration which is a solution of \eqref{eq:rho_of_theta} and $\xi$ can be either the function $\sin(\cdot)$ or $\cos(\cdot)$. Since all curves for all $J$ pass through the point $(1/2,\pi/4)$, we can analyze these equations at $\theta^*=\pi/4$ in order to specify $J_c$. Indeed, $J_c$ is the value where the function $r(\theta)$ has a saddle point, meaning that both its first and second derivative evaluated at this point vanish. The first derivative takes the form
\begin{equation}
   r'(\theta)|_{\theta=\frac{\pi}{4}}=1-\frac{m'\left(\frac{1}{\sqrt{2}}\right)}{\sqrt{2} m\left(\frac{1}{\sqrt{2}}\right)}\overset{!}{=}0\,.
   \label{eq:First_derivative}
\end{equation}
Importantly, note that the arguments of $1/\sqrt{2}$ arising in \eqref{eq:First_derivative} come from evaluating $\sin(\pi/4)=\cos(\pi/4)=1/\sqrt{2}$. When imposing \eqref{eq:First_derivative} for an explicit potential $m(\xi)$, one should first compute its derivative for a generic argument $\xi$ and subsequently evaluate it at $\xi=1/\sqrt{2}$. Indeed, it is only after the insertion of an explicit potential $m(\xi)$ that the parameter $J$ enters, and it is precisely the condition \eqref{eq:First_derivative} which singles out the value of $J_c$ for that specific potential. Since we are evaluating functions at a specific value, the constraint \eqref{eq:First_derivative} is no longer a partial differential equation but rather a simple algebraic equation with solution $m'(1/\sqrt{2})=\sqrt{2}m(1/\sqrt{2})$. The second derivative is given by
\begin{equation}
    r(\theta)''|_{\theta=\pi/4}=\frac{m'\left(\frac{1}{\sqrt{2}}\right)^2}{m\left(\frac{1}{\sqrt{2}}\right)^2}-\frac{2 \sqrt{2} m'\left(\frac{1}{\sqrt{2}}\right)}{m\left(\frac{1}{\sqrt{2}}\right)}+2\overset{!}{=}0\,.
    \label{eq:Second_derivative}
\end{equation}
Indeed, we see that the equation \eqref{eq:Second_derivative} is solved automatically by the solution of \eqref{eq:First_derivative} for a generic function $m(\xi)$ when inserting the values $r=1/2$ and $\theta=\pi/4$. Therefore, we conclude that when the potential obeys $m'(1/\sqrt{2})=\sqrt{2}m(1/\sqrt{2})$ as in \eqref{eq:First_derivative}, this condition singles out a specific value of $J$, namely $J_c$, for which the solution $r(\theta)$ has a saddle point at $\theta=\pi/4$.\\
Let us take a look at explicit potential examples. For the exponential potential we have $m(\xi)=-e^{J \xi}/2$, the condition \eqref{eq:First_derivative} is solved straightforwardly and yields $J_c=\sqrt{2}$. We can also consider truncations of the series expansion of the exponential, i.e potentials of the form
\begin{equation}
    m(\xi)=-\frac{1}{2}\sum_{k=0}^n\frac{1}{k!}(J \xi)^k\,.
    \label{eq:Polynomial_Potential}
\end{equation}
These truncations describe the cases where we restrict the amount of interaction terms in the original Hamiltonian. We can evaluate the condition \eqref{eq:First_derivative} to obtain
\begin{equation}
\sum_{k=0}^n\frac{k-1}{k!}\frac{1}{(\sqrt{2})^{k-1}}J^k=0\,.
\label{SMeq:J_crit_defining_eq}
\end{equation}
We thus see that determining the value of $J_c$ boils down to finding the roots of an order $n$ polynomial with constant coefficients. Therefore, the critical point $J_c$ can be found analytically for all potential of the form given in \eqref{eq:Polynomial_Potential}. Importantly, the multi-valuedness of the solutions to the SC equation is an indicator that the system undergoes a phase transition, as we identify in more detail in the next section. More precisely, this phase transition can be detected by computing the free energy on the SC solutions, as we present in Fig.~\ref{fig:Multipanel_Free_Energy} for different choices of the potential.

\subsubsection{Free energy and phase transition}

We now give a derivation of the free energy $F$ for the four-site chain with generic potentials. This derivation follows the same steps as that of Section \enquote{Free energy from path integral} for the two-site chain, with the main differences being the effective action and the physical parameters entering the problem. In the case of the four-site chain, the last term of the effective action \eqref{SMeq:eff_action} reads
\begin{equation}
\begin{split}
    -\frac{1}{2} \sum_{x,y} \int d\tau h_{xy}(-2\mathrm{i}G_{xy}(0))
    &=-\textrm{Vol}(\tau)\big(2 h_a(-2\mathrm{i}G_{12}(0))+h_b(-2\mathrm{i}G_{23}(0))\big)\,\\
    &=-\textrm{Vol}(\tau)\mu_b\big( 2r \int^{X_{12}} d\xi m(\xi) + \int^{X_{23}} d\xi m(\xi)\big)\equiv S_h[G](r)\,,
\end{split}
\label{SMeq:eff_action_four_site}
\end{equation}
where we have exploited the anti-symmetry of $G$, the property $h_{x,x+1}(\xi)=h_{x+1,x}(-\xi)$, and the fact that $G_{12}(0)=G_{34}(0)$. We have also written the potential in the form used in previous sections, $h_{a,b}(\xi)=\mu_{a,b}\int^{\xi} d\xi' m(\xi')$, where the two integrals are over the same function $m$ but evaluated at different values $\xi=X_{12}\equiv -2\mathrm{i}G_{12}(0)$ and $\xi=X_{23}\equiv -2\mathrm{i}G_{23}(0)$, in a slight but necessary abuse of notation. Further, note that the integrals over Euclidean time appearing in the effective action \eqref{SMeq:eff_action} are trivial and give only a constant contribution, since the self-energy \eqref{SMeq:self_energy_four_sites} contains a Dirac delta in $\tau$, therefore leaving only $G(0)$ present in the action. More precisely, there is a constant (divergent) contribution which we denote as $\int d\tau\equiv \textrm{Vol}(\tau)$ coming from the integrals in the action. Finally, we have extracted a factor of $\mu_b$ such as to identify the hopping ratio $r$ in the first term of the second line in \eqref{SMeq:eff_action_four_site}. We can now split the total action into three terms $-S_E/N\equiv S_{\textrm{Pf}}[\Sigma]+S_G[G,\Sigma]+S_h[G](r)$. In a similar fashion as for the two-site chain, we avoid the evaluation of the Pfaffian by performing a derivative with respect to the dimensionless hopping ratio $r$,
\begin{equation}
    \begin{split}
        r\frac{\partial}{\partial r}\left(-\frac{S_E}{N}\right)
        &=r \frac{\partial \Sigma}{\partial r}\left(\frac{\delta S_{\textrm{Pf}}}{\delta \Sigma}+\frac{\delta S_{G}}{\delta \Sigma}\right)+r \frac{\partial G}{\partial r}\left(\frac{\delta S_{G}}{\delta G}+\frac{\delta S_{h}}{\delta G}\right)+r\left(\frac{\partial S_h}{\partial r}\right)\,.
    \end{split}
\end{equation}
Evaluating this expression on-shell, the first two terms vanish because the fields $G,\Sigma$ obey the equations of motion. The only non-vanishing contribution comes from the explicit dependence on the hopping ratio, and thus we have
\begin{equation}
    \frac{\partial}{\partial r}\left(-\frac{S_E}{N}\right)_{|_{\textrm{\tiny on-shell}}}=\frac{\partial}{\partial r}\left(S_h[G](r)\right)_{|_{\textrm{\tiny on-shell}}}=-2\textrm{Vol}(\tau)\mu_b \int^{X_{12}} d\xi m(\xi)\,,
\end{equation}
where once again the SC condition imposes $X_{12}=-2\mathrm{i}G_{12}(0)$, similarly to the two-site case. The difference is than now $G_{12}(0)$ is the two-point function of the four-site chain, cf.~\eqref{eqSM:4site_Greenfunctions} at $\tau=0$ or \eqmaintext{10}.\\
In order to obtain the free energy, one would need to integrate back over $r$, albeit in a suitable parametrization. A useful one is in terms of the parameter $\theta$ appearing in the SC equation, which gives $-2\mathrm{i}G_{12}(0)=\sin\theta$ and for which $r$ is itself dependent on $\theta$ and the interaction potential via \eqref{eq:rho_of_theta}. Implementing this change of variables, the general expression for the free energy reads
\begin{equation}
    \frac{\textrm{Vol}(\tau)F}{N}=2 \textrm{Vol}(\tau)\mu_b \int_{r_0}^{r} dr'\left[\int^{X_{12}}d\xi m(\xi)\right]_{X_{12}=\sin\theta}= 2 \textrm{Vol}(\tau)\mu_b \int_{\theta_0}^{\theta} d\theta' \left[\frac{\partial r(\theta'')}{\partial \theta''}\right]_{\theta''=\theta'}\left[\int^{X_{12}}d\xi m(\xi)\right]_{X_{12}=\sin\theta}\,.
    \label{SMeq:General_Onshell_Action_Four_Sites}
\end{equation}
This expression is somewhat reminiscent of the one obtained in \eqref{eq:General_Onshell_Action_FiniteT} for the two-site case. In the latter, due to the simple nature of the two-site system, the integral could be evaluated without explicit knowledge of the functions $h$ or $m$. For the four-site chain this is no longer the case, and the integral \eqref{SMeq:General_Onshell_Action_Four_Sites} can only be evaluated explicitly after choosing a specific form of the interaction potential. For the case of the exponential potential discussed in the main text,  we have $m(\xi)=-e^{J\xi}/2$, and we fix its integral to give the interaction potential mentioned in the main text, i.e. $h_x(\xi)=\mu_x\int d\xi' m(\xi')=\frac{\mu_x}{2J}(1-e^{J\xi})$. Inserting these expressions into \eqref{SMeq:General_Onshell_Action_Four_Sites}, we obtain the free energy of the four-site chain
\begin{equation}
\frac{F}{N}=\frac{\mu_b}{2J}
\left[\tan (\theta ) e^{J  (\cos (\theta )-\sin (\theta ))}-(\tan (\theta )+1) e^{J  \cos (\theta )}+e^{J}\right]\,,
\label{SMeq:free_energy_aba}
\end{equation}
where the last term $e^{J}$ results from evaluating the integration constant at $\theta=0$,
\begin{equation}
    \left[\tan(\theta_0)e^{J (\cos(\theta_0)-\sin(\theta_0))}-(1+\tan(\theta_0)e^{J \cos(\theta_0)})\right]\Big|_{\theta_0=0}=-e^{J}\,.
\end{equation}
This is the free energy plotted in Fig.~1 of the main text for different values of the interaction strength $J$. Without the loss of generality, we have set $\mu_b=1$ for all the plots. Recall that the free energy is to be evaluated on solutions $\theta$ to the SC equation \eqref{eq:rho_of_theta}, which can be multi-valued for $J>J_c$ as discussed in the previous sections of this SM. This multi-valuedness is then also present in the plots for the free energy. The SC solutions which minimize the free energy \eqref{SMeq:free_energy_aba} are the thermodynamically dominant ones, and it can be seen from Fig.\,\ref{SMfig:tantheta_plot} that these exhibit a discontinuity at $r=1/2$ for $J>J_c$. This leads to a non-analyticity of the free energy \eqref{SMeq:free_energy_aba}, signaling the phase transition. We can identify two phases, one for $r>1/2$ and the other for $r<1/2$, and the function $(\tan(\theta))(r)$ reveals the character of each of the phases. In the phase of $r<1/2$, $\tan\theta$, and thus $\theta$, is small, and therefore the correlations that dominate the system according to the Green's function given in \eqmaintext{10} are $G_{23}$ and $G_{14}$. In the phase of $r>1/2$, $\tan \theta$ is large, implying that $\theta$ approaches $\pi/2$. Again, comparing with the Green's function given in \eqmaintext{10}, this indicates that in this phase the correlations $G_{12}=G_{34}$ are the dominant ones. This correlation structure is shown in the insets of Fig.~1 of the main text for the limits $r\to 0$ and $r\to \infty$.\\
\begin{figure}[!t]

\subfloat[\label{multifig:Linear_Pot}]{%
  \includegraphics[width=0.5\columnwidth]{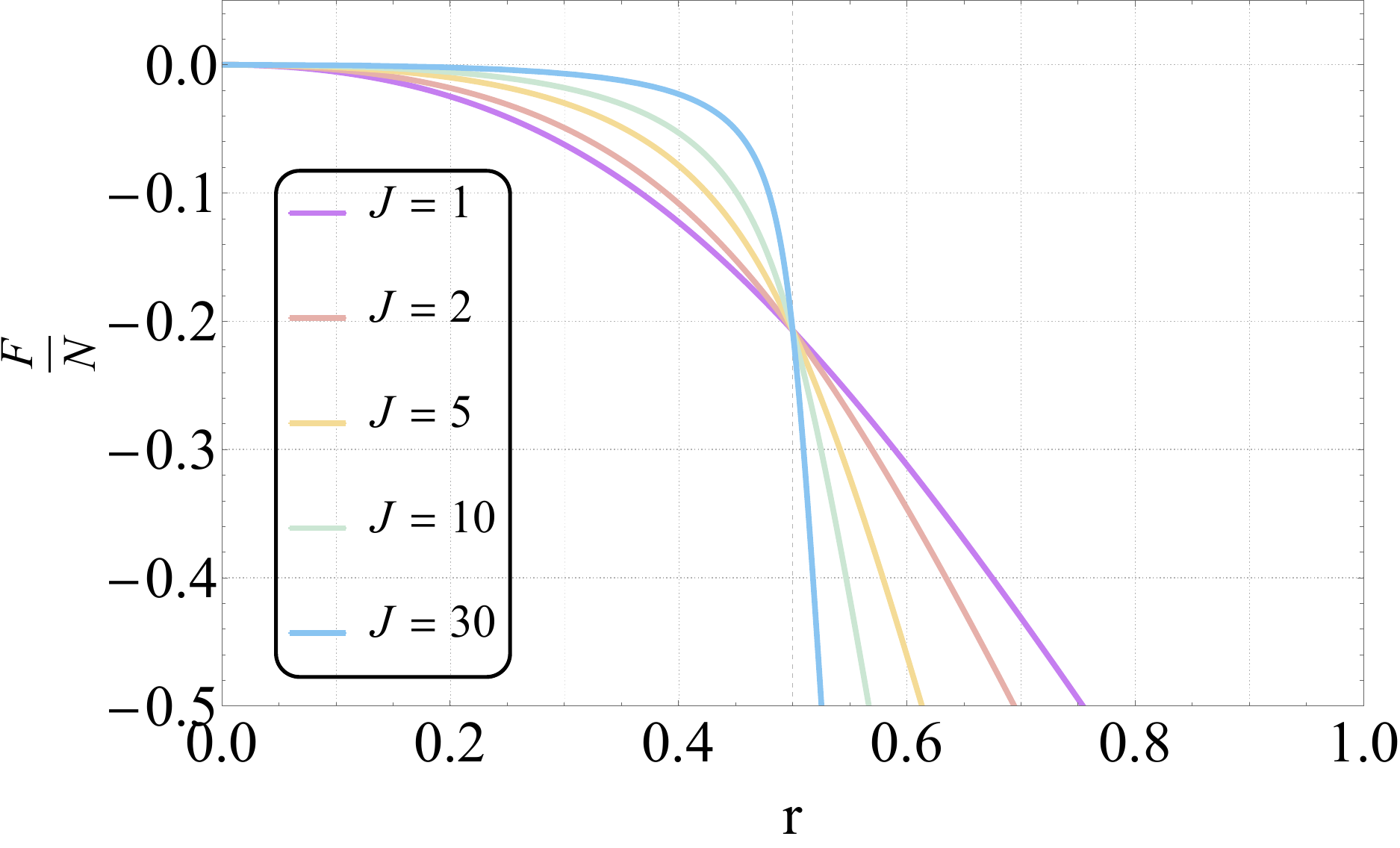}%
}\hfill
\subfloat[\label{multifig:Quadratic_Pot}]{%
  \includegraphics[width=0.5\columnwidth]{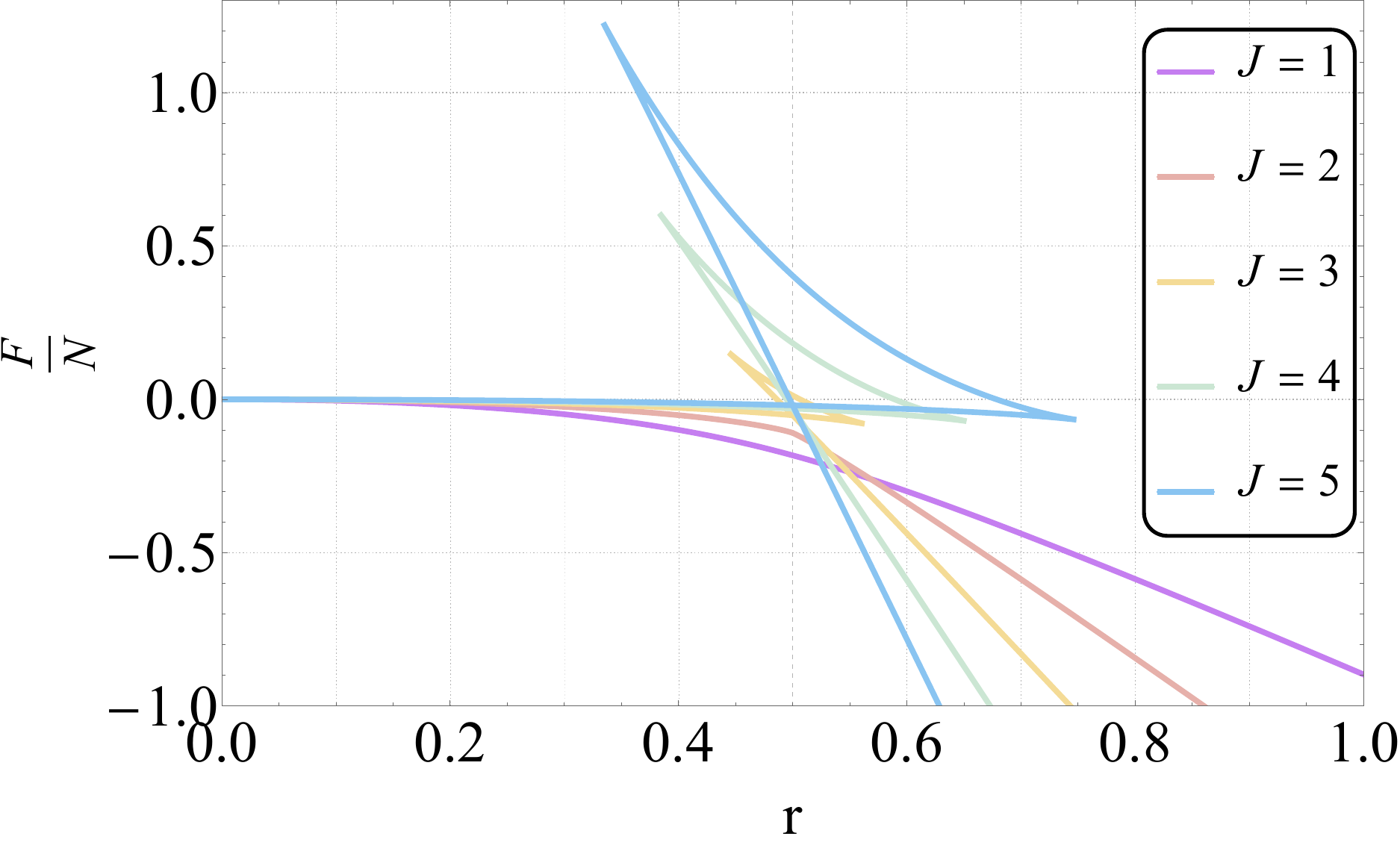}%
}

\subfloat[\label{multifig:Cubic_Pot}]{%
  \includegraphics[width=0.5\columnwidth]{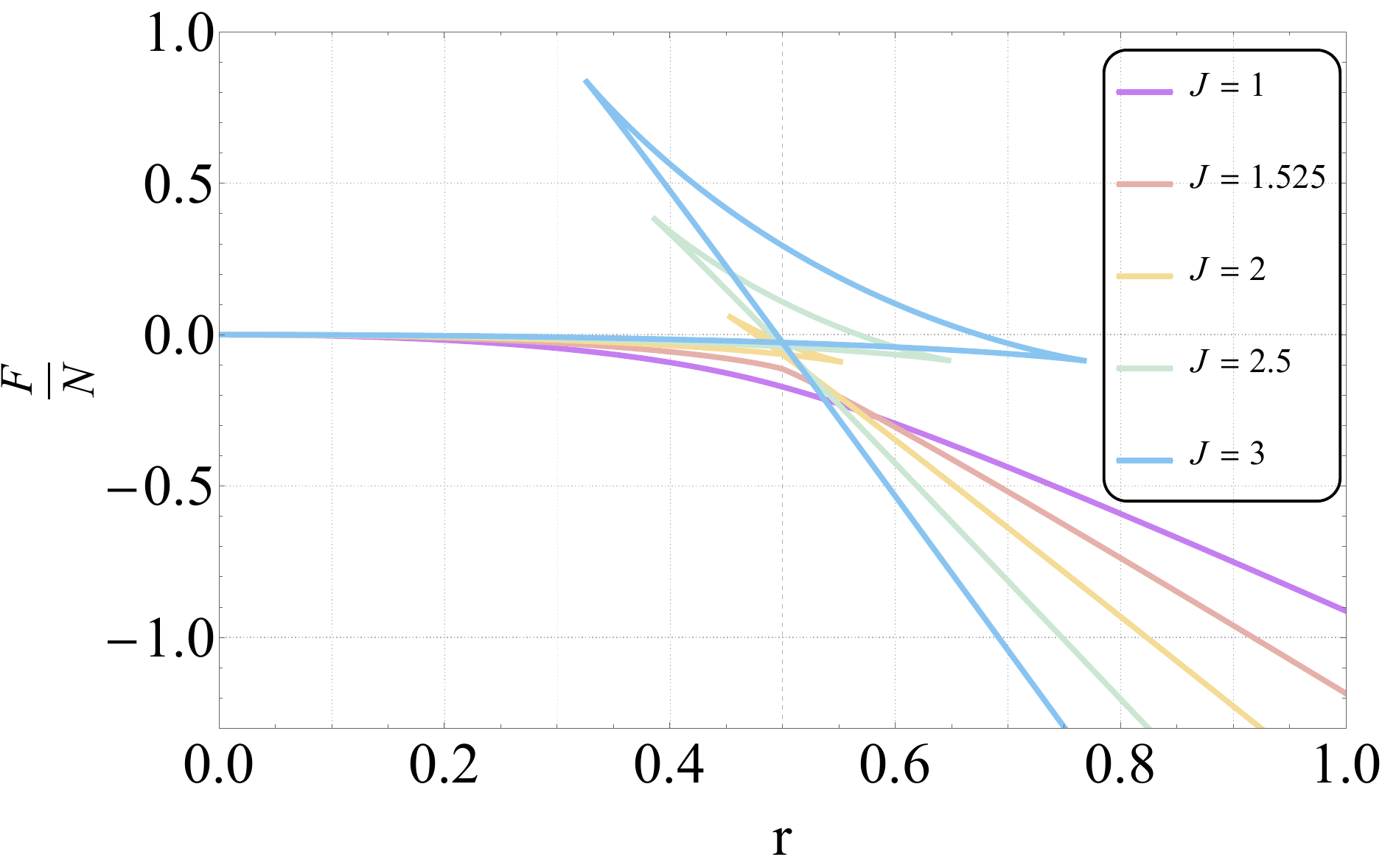}%
}\hfill
\subfloat[\label{multifig:Quartic_Pot}]{%
  \includegraphics[width=0.5\columnwidth]{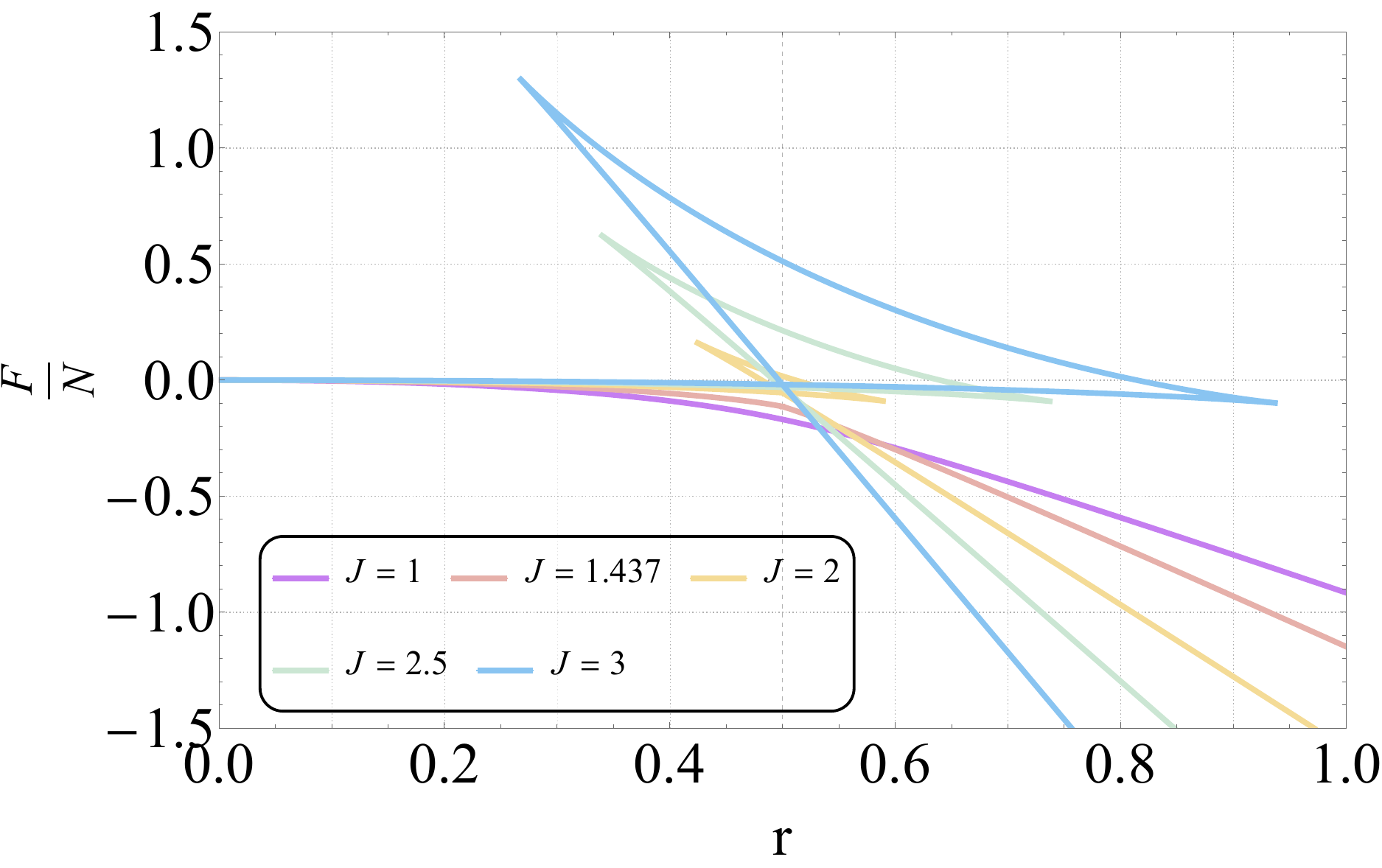}%
}\hfill
\caption{Free energy \eqref{SMeq:General_Onshell_Action_Four_Sites} for linear (\ref{multifig:Linear_Pot}), quadratic (\ref{multifig:Quadratic_Pot}), cubic (\ref{multifig:Cubic_Pot}), and quartic (\ref{multifig:Quartic_Pot}) interaction potential functions $m$ in the form \eqref{eq:Polynomial_Potential}. Their respective critical interaction strengths are $J_c=\infty$ for the linear, $J_c=2$ for the quadratic, $J_c\sim 1.525$ for the cubic, and $J_c\sim 1.437$ for the quartic potential.}
\label{fig:Multipanel_Free_Energy}
\end{figure}
For completeness, we evaluate the free energy  \eqref{SMeq:General_Onshell_Action_Four_Sites} for some examples of truncated polynomial potentials $m$ of the form given in \eqref{eq:Polynomial_Potential} and for different values of the interaction strength $J$. In particular, we consider linear, quadratic, cubic and quartic potential functions $m$. The resulting free energy computed via \eqref{SMeq:General_Onshell_Action_Four_Sites} is shown in the panels of Fig.\,\ref{fig:Multipanel_Free_Energy}. The critical values $J_c$, if existent, have been obtained through \eqref{SMeq:J_crit_defining_eq}, and the free energy at these critical values is always displayed, together with values of $J$ below and above the critical one. The linear case is the only one not exhibiting a phase transition, given that its critical interaction strength is $J_c=\infty$. The critical values for the truncated polynomial potentials approach the critical value of the exponential potential when the order is increased. This is expected since the series expansion \eqref{eq:Polynomial_Potential} approaches that of the exponential when the order $n\to\infty$. For the other three potentials shown as an example here, the system undergoes a phase transition at $r=1/2$ for $J>J_c$, where the two phases are characterized by their correlation structure, as explained above for the case of the exponential potential, cf.~Fig.~\ref{SMfig:tantheta_plot}.

\subsection{von Neumann algebras}\label{apx:vN_algebras}

\subsubsection{Modular operator for two-site subregions without interaction}

In what follows, we consider the four-site chain with Hamiltonian (1) of the main text and potential given by $h_{x,x+1}(\xi)=h_{x+1,x}(-\xi)=-\mu_x/2$, with $\mu_1=\mu_3=\mu_a$, $\mu_2=\mu_b$. This choice corresponds to having a free model, for which explicit results for the reduced density matrix are known. This case therefore provides a rich playground to gain insights into the entanglement structure in general, and the properties of the underlying operator algebras in particular.\\
In this free case, the ground state factorizes into contributions from the different species (labeled by $j$) of Majorana fermions and, due to this fact, the model can be solved for any value of $N$.  Because of the quadratic nature of the model, the reduced density matrix of a two-site subsystem $A$ has the form
\be
\label{eq:EH_largeN_general}
\rho_A=\frac{e^{-\sum_{x,y=1}^2 K_{xy}\sum_{j=1}^N\psi^j_x\psi^j_y}}{Z}
\,,    
\ee
where the two sites in the subsystem are labeled by 1 and 2, the constant $Z$ ensures the normalization $\mathrm{Tr}\rho_A=1$ and the $2\times 2$ matrix $K_{xy}$ is the only unknown to determine $\rho_A$. Let us notice that $(\psi_x^j)^2=1/2$ and the constant resulting from these terms can be reabsorbed in the normalization, making the expression of $\rho_A$ simpler, i.e.
\begin{equation}
\label{eqSM:RDM_2sites_v0}
  \rho_A=\frac{e^{- (K_{21}-K_{12})\sum_{j=1}^N\psi^j_2\psi^j_1}}{Z}
  = \frac{e^{- \mathrm{i}2\kappa\sum_{j=1}^N\psi^j_2\psi^j_1}}{Z} 
  \,,
\end{equation}
where $K_{12}=-K_{21}\equiv-\mathrm{i}\kappa$.  Given that $[\psi^j_2\psi^j_1,\psi^i_2\psi^i_1]=0$, $\rho_A$ factorizes into $N$ terms, each associated with a different Majorana species, and so does the partition function. The $N$ contributions to $Z$ have the same expression, leading to
\be
Z=\left(2\cosh\kappa\right)^N.
\ee
For free fermionic systems, $K$ (and therefore $\kappa$) is uniquely fixed in terms of the two-point functions within the subsystem $A$ \cite{IngoPeschel_2003}. Given the expression (\ref{eqSM:RDM_2sites_v0}), we can relate $\kappa$ to the two-point function within $A$ by
\begin{equation}
    G_{21}(0)\equiv\frac{1}{N}\sum_{j=1}^N\langle\psi^j_2\psi^j_1\rangle=\frac{\mathrm{i}}{2}\tanh{\kappa}\,.
    \label{eq:kappafromGreen}
\end{equation}
From the knowledge of $G_{21}(0)$ (see \eqref{eqSM:4site_Greenfunctions} at $\tau=0$ or (10) in the main text), we can determine the expression of $\kappa$ , and therefore $\rho_A$, in terms of the parameters of the model. In particular, we find
\begin{equation}
\label{eqSM:kappa_free}
    \kappa=\frac{1}{4} \log \left(\frac{\cos (2 \theta )+4 \sin (\theta )-3}{\cos (2 \theta )-4 \sin (\theta )-3}\right)\,,
\end{equation}
where in the free case $\theta$ is related to the hopping ratio as $\frac{\mu_a}{\mu_b}=\frac12\tan\theta$.
A straightforward computation allows to compute the entanglement entropy $S_A=-\textrm{Tr}\left(\rho_A\log \rho_A\right)$, leading to 
\begin{equation}
\label{eqSM:EE_2sites}
    S_A=N\left[\log\left(1+e^{2\kappa}\right)-\kappa\left(1+\tanh{\kappa}\right)\right]\,.
\end{equation}
Let us stress that, in the interacting case, the form (\ref{eqSM:RDM_2sites_v0}) of the reduced density matrix is no longer valid. Nevertheless, the expression (\ref{eqSM:EE_2sites}) for the entanglement entropy in the large $N$ limit with $\kappa$ in (\ref{eqSM:kappa_free}) holds also for the interacting case, as long as $\theta$ is a solution of the SC condition \eqref{eq:rho_of_theta} or \eqmaintext{11}. Crucially, in the free case, \eqref{eqSM:EE_2sites} holds for any value (finite and infinite) of $N$, while in the interacting case it is reliable only at leading order in large $N$.\\
As remarked in the main text, a useful object for classifying the von Neumann operator algebras associated with our model in the large $N$ limit is the modular operator
\begin{equation}
   \Delta=\lim_{N\to\infty}\rho_A\otimes \rho^{-1}_{\bar{A}} \,,
\end{equation}
where $\rho_{\bar{A}}$ is the reduced density matrix of the complement subsystem, which in the considered four-site chain has the same form as  $\rho_A$. In the free case, the ground state of the model factorizes into the (infinite) tensor product of vectors in finite-dimensional Hilbert spaces. This structure allows to determine the type of the operator algebras acting on the Hilbert space of the system by studying the spectrum of $\Delta$ \cite{vNeumann1939,Powers1967,Araki1968,Takesaki1979}. Notice that for this particular case, the classification using the modular operator is equivalent to the definitions given in the main text in terms of trace functionals. Combining the the expressions of $\rho_A$  and $\rho_{\bar{A}}$, we obtain
\begin{equation}
 \Delta=\lim_{N\to\infty}\bigotimes_{j=1}^N
    \begin{pmatrix}
        1&0&0&0\\
        0&e^{-2\kappa}&0&0\\
        0&0&e^{2\kappa}&0\\
        0&0&0&1
    \end{pmatrix}
    \end{equation}
which leads to the spectrum
\begin{equation}
    \textrm{Spec}(\Delta)=\lim_{N\to\infty}\left\{
    e^{2n\kappa}
    \right\}_{n=-N}^{N}\,,
\end{equation}
where we are not taking into account the degeneracies. Let us discuss the behavior of the spectrum in the various regimes of $\kappa$. When $\kappa=0$, all the eigenvalues of $\Delta$ are 1, leading to an underlying operator algebra of type II$_1$.  For $-\infty<\kappa<0$, the spectrum is of the form $\lambda^n$, with $n\in\mathbb{Z}$ and $0<\lambda\equiv e^{2\kappa}<1$. This property of the spectrum of the modular operator characterizes operator algebras of the so-called type III$_\lambda$. Finally, when $\kappa\to-\infty$, all the factors in (\ref{eqSM:RDM_2sites_v0}) become projectors and the resulting entanglement entropy vanishes, yielding a type I$_\infty$ operator algebra.\\
This analysis holds for any two-site subsystem in the four-site chain and therefore includes the case considered in the main text where $A$ is the left half of the chain. These considerations lead to the results reported in the dedicated section of the main text. Indeed, if one relates $\kappa$ to the correlations $X$ within the subsystem $A$ as $X=-\tanh \kappa$, we have that $\kappa\to 0$ implies $X\to 0$, $-\infty<\kappa<0$ implies $0<X<1$ and $\kappa\to -\infty$ implies $X\to 1$. Thus, we retrieve the algebra typification in the various regimes of the phase diagram summarized in Fig.~2 of the main text.

\subsubsection{Large \texorpdfstring{$J$}{J} limit}

We provide details on our analysis of the limit $J\to\infty$, focusing in particular on the case of an exponential interaction potential. Based on Fig.~2 of the main text, it appears that the behavior of the entanglement entropy density $S_A/N$ is that of a step function when $J\to\infty$, abruptly going from entanglement $S_A/N=\log 2$ to zero entanglement $S_A/N=0$ at $r\equiv\mu_a/\mu_b=1/2$. We now make these considerations more precise by analyzing the SC constraint given in \eqmaintext{11} in this limit.\\
We start by recalling that in order to evaluate the entanglement entropy, we need to find the solutions to the SC equation (cf. \eqref{eq:Self_Cons_Eq} and \eqmaintext{11})
\begin{equation}
    r e^{J(\sin{\theta}-\cos{\theta})}=\frac{1}{2}\tan{\theta}\,,
    \label{SMeq:SC_aba_second_version}
\end{equation}
For $J>J_c=\sqrt{2}$, which is clearly the case if $J\to\infty$, there are three possible solutions, the thermodynamically dominant one being that which minimizes the free energy functional \eqref{SMeq:free_energy_aba}. For the four-site chain at zero temperature, we know that the SC equation \eqref{SMeq:SC_aba_second_version} has a fixed point which is independent of $J$, namely $\theta^*=\pi/4$ (cf. discussion below \eqref{eq:Self_Cons_Eq}). This fixed point also imposes that $r=1/2$.\\
Consider now the two cases of slight deviations from this fixed point, namely $\theta_{\pm}=\theta^*\pm\epsilon$, with $\epsilon\ll 1$. Note that these deviations automatically imply that we restrict ourselves to values of $r$ slightly above or below $1/2$, respectively. For $\theta_+$, we have that the exponent in the SC equation is positive, i.e. $(\sin\theta-\cos\theta)>0$. This implies that $e^{J(\sin{\theta}-\cos{\theta})}\to\infty$ as $J\to\infty$. For self-consistency, the r.h.s. of the SC equation \eqref{SMeq:SC_aba_second_version} must therefore also diverge, even though it has no explicit $J$-dependence. From $\tan\theta\to\infty$ we extract that the solution to the SC must tend to $\theta\to\pi/2$. Using the relation \eqref{eqSM:kappa_free} (recall that the latter holds both for free and interacting cases), this implies that $\kappa\to-\infty$, which in turn yields $X\to 1$ after recalling that $X=-\tanh{\kappa}$ by virtue of \eqref{eq:kappafromGreen}. Thus, the corresponding entanglement entropy vanishes $S_A\to 0$ according to \eqref{SMeq:entropy_function} and \eqmaintext{9}. Analogously, for the case of the deviation $\theta_-$, we have the opposite behavior in which $e^{J(\sin{\theta}-\cos{\theta})}\to 0$ as $J\to\infty$, implying that the self-consistent solution tends to $\theta\to 0$. Again using \eqref{eqSM:kappa_free} and \eqref{SMeq:entropy_function} (or \eqmaintext{9}), we see that this implies $\kappa\to 0$ ($X\to0$) and therefore entanglement $S_A/N\to\log2$. Thus, we have shown that in the limit of $J\to \infty$, the entanglement entropy is maximal (at leading order for $N\to\infty$) in the whole regime $0\leq r<1/2$, while it is vanishing for $r>1/2$, based on the behavior of the solutions to the SC equation.\\
Remarkably, this phenomenon is unique to the class of potentials with exponential behavior. This is because for any potentials given by a general truncated series expansion of the form \eqref{eq:Polynomial_Potential}, the behavior of the l.h.s. of \eqref{SMeq:SC_aba_second_version} when $J\to\infty$ does not impose a constraint on the r.h.s.. More precisely, for exponential potentials we were left with the function $e^{J(\sin{\theta}-\cos{\theta})}$, which has a clear-cut limiting behavior as $J\to\infty$ based on the value of $\theta$. In contrast, for any truncated potential, the limit $J\to\infty$ would make the highest-power term dominate, such that 
\begin{equation}
    \frac{m(\sin\theta)}{m(\cos\theta)}=\frac{\sum_{k=0}^n\frac{J^k}{k!} (\sin\theta)^k}{\sum_{k=0}^n\frac{J^k}{k!} (\cos\theta)^k}\overset{J\to\infty}{\sim}(\tan\theta)^n>1\,,\,\quad\forall\,\,\theta\in[0,\frac{\pi}{2})\,.
\end{equation}
The SC equation \eqref{SMeq:SC_aba_second_version} in this case takes on the form
\begin{equation}
    r (\tan\theta)^n=\frac{1}{2}\tan\theta\,,
\end{equation}
which indeed has uniquely defined solutions given by the smooth function $\theta=\tanh^{-1}((2r)^{\frac{1}{1-n}})$ throughout the whole range of $\theta$. Therefore, we see that for truncated polynomial interaction potentials, the SC solutions exhibit a smooth behavior. Thus, the entanglement entropy does no longer have a discontinuity.\\
We further report evidence as to how the limit $J\to\infty$ leads to a Heaviside step function $\Theta$ centered at $r=1/2$ for both the correlations $X=\Theta(r-1/2)$ and correspondingly the entanglement entropy $S_A/N=\log 2 \Theta(1/2-r)$ for the case of an exponential interaction potential. Consider the deviation $\theta_+$. We know from the above considerations that the self-consistent solution should be given in this regime by $\theta=\pi/2$. Let us therefore expand the SC equation with respect to a small deviation $\varepsilon\ll1$ from this point. At first order, we have
\begin{equation}
    r=e^{-J}\left(-\frac{1}{2 \left(\theta -\frac{\pi }{2}\right)}+\frac{J}{2}\right)\,,
\end{equation}
which can be solved to obtain a direct relation for $\theta$ as a function of $r$ and the interaction strength $J$
\begin{equation}
    \theta \left(r,J\right)=\frac{-2 \pi  e^{J } r +\pi  J +2}{2 \left(J -2 e^{J } r \right)}\,.
\end{equation}
Exploiting the relation \eqref{eqSM:kappa_free} between $\kappa$ and $\theta$, we can immediately find the function $X(\kappa(\theta))$ as well by virtue of \eqref{eq:kappafromGreen}. By inserting this relation into the entanglement entropy given in \eqref{SMeq:entropy_function} we find its behavior at large $J$ and for $r>1/2$
\begin{equation}
    \frac{S_A}{N}=-2\log \left(\cos\left(\frac{\zeta}{2}\right)\right)+\left(1-\cos \left(\zeta\right)\right) \tanh^{-1}\left(\cos \left(\zeta\right)\right)\,,
    \label{eq:EE_expanded_large_lambda}
\end{equation}
where we have defined the parameter
\begin{equation}
    \zeta\equiv\frac{1}{2 e^{J } r-J }\,.
    \label{eq:parameter_zeta}
\end{equation}
Note that this expression is not necessarily an exact solution of the SC, but rather an approximation which is very good in the regimes of $r$ far away from $1/2$. Around $1/2$, the validity of the approximation increases as $J$ gets larger. The expression \eqref{eq:parameter_zeta} formally allows us to set $r=1/2$ and then take the limit $J\to\infty$. However, since we are working with an approximation as mentioned above, it is more proper to leave the value of the hopping ratio $r$ free and analyze the overall behavior of the expression \eqref{eq:EE_expanded_large_lambda} when $J\to\infty$. It is straightforward to see that 
\begin{equation}
\lim_{J\to\infty}\zeta=0\,,\quad \forall\;\;r\,,
\end{equation}
and therefore the first term in \eqref{eq:EE_expanded_large_lambda} vanishes in this limit. The second term in \eqref{eq:EE_expanded_large_lambda}, however, leads to an indeterminate expression of the type $0\cdot\infty$. Writing $f(\zeta)/(1/g(\zeta))$, with $f(\zeta)=\tanh^{-1}{(\cos(\zeta))}$ and $g(\zeta)=(1-\cos(\zeta))$, we transform the expression to an indeterminate form of type $\frac{\infty}{\infty}$ and apply L'Hôpital's rule, obtaining
\begin{equation}
    \lim_{J\to\infty}f(\zeta)g(\zeta)=\lim_{J\to\infty}\frac{f(\zeta)}{1/g(\zeta)}=\lim_{J\to\infty}\frac{\partial_{J}f(\zeta)}{\partial_{J}(1/g(\zeta))}=\lim_{J\to\infty}\tan^2(\zeta)=0\,.
    \label{eq:EE_large_lambda_limit}
\end{equation}
We thus see that the value of the entanglement entropy for a value of $r$ infinitesimally larger than $1/2$ tends to zero in the large $J$ limit. Therefore, the entropy as a whole tends to zero for any $r>1/2$ when $J\to\infty$.\\
In fact, the exact same analysis can be done for the other relevant value of the SC equation, namely the point towards which $\theta$ is pushed when slightly deviating from $r=1/2$ by $-\epsilon$. In this case, from the considerations of the previous paragraph, the solution to the SC equation is $\theta=0$. We can again expand the SC equation around this point and insert the resulting dependence $\theta(r,J)$ into the formula for the entanglement entropy, obtaining an expression very similar to \eqref{eq:EE_expanded_large_lambda}
\begin{equation}
    \frac{S_A}{N}=\log \left(\frac{2}{\sin \left(\tilde{\zeta}\right)+1}\right)+\left(1-\sin \left(\tilde{\zeta}\right)\right) \tanh^{-1}\left(\sin \left(\tilde{\zeta}\right)\right)\,,
    \label{eq:EE_expansion_leftregime}
\end{equation}
with the parameter
\begin{equation}
    \tilde{\zeta}=2\,r\,e^{-J }\,.
\end{equation}
The limit of $J\to\infty$ of this expression can be taken straightforwardly since $\lim_{J\to\infty}\Tilde{\zeta}=0$ and there are no indeterminate expressions, finding 
\begin{equation}
    \lim_{J\to\infty}\frac{S_A}{N}=\log 2.
\end{equation}
Once again, we see that the value of the entropy density for a value of $r$ infinitesimally smaller than $1/2$ tends to $\log 2$. The entropy density as a whole thus tends to this value for all $r<1/2$ when $J\to\infty$.

\subsection{Periodic chain}\label{apx:periodic_ab_chain}

We consider the closed periodic chains discussed in the dedicated section of the main text. We assume to have a chain with $L$ sites described by the Hamiltonian (1) in the main text with a staggered interaction of the form (13). In our conventions, $x=0,\dots, L-1$ labels the site of the chain and the periodic boundary conditions impose the identification $x\sim L+x$ for this variable. This model exhibits translational invariance with respect to cells of adjacent sites interacting by $h_b$. More precisely, we can introduce $r=0,\dots,\frac{L}{2}-1$, such that $x=2r+s$, with $s\in\{0,\,1\}$. The chain is then invariant with respect to the translation of the index $r$. Given $x=2r_1+s_1$ and $y=2r_2+s_2$, $h_{xy}\equiv h_{r_1r_2}^{s_1 s_2}$ and we can rewrite the derivative of \eqmaintext{13} as
\begin{equation}
    {h'}_{r_1r_2}^{s_1 s_2}=M_b\delta_{r_1 r_2}\epsilon^{s_1s_2}
    +M_a\left(\delta_{r_1-r_2,-1}\delta_{1,s_1}\delta_{0,s_2}
    -\delta_{r_1-r_2,1}\delta_{0,s_1}\delta_{1,s_2}\right)\,,
\end{equation}
where $\epsilon^{s_1s_2}$ is the totally anti-symmetric symbol such that $\epsilon^{0 1}=1$ and we have further defined
\begin{equation}
    M_a\equiv h'_a\left(-2 \mathrm{i}G^{1,0}_{r,r+1}(0)\right)\,,
\qquad
 M_b\equiv h'_b\left(-2 \mathrm{i}G^{0,1}_{r,r}(0)\right)\,,
 \label{eqSM:Ma_Mb_periodic}
\end{equation}
exploiting the fact that the components of ${h'}_{r_1r_2}^{s_1 s_2}$ depend on the indices $r_1$ and $r_2$ only through their differences due to the aforementioned translational invariance.\\
For convenience, we also denote the spatial dependence of $G_{xy}$ in terms of $r_1$, $r_2$, $s_1$ and $s_2$. Given that ${h'}_{r_1r_2}^{s_1 s_2}$ depends only on $r_1-r_2\equiv \delta r$ and on the two indices $s_1$ and $s_2$, it is particularly convenient to write the entries of ${h'}_{r_1r_2}^{s_1 s_2}$ uniquely as function of $\delta r$ and collect them into $2\times 2$ matrices in the indices $s_1$ and $s_2$. We thus have
\begin{equation}
\label{eqSM:potentialMatrix}
    {h'}_{r_1r_2}^{s_1 s_2}=\begin{pmatrix}
        0   &  M_b\delta_{\delta r,0}-M_a\delta_{\delta r,1} \\
        -M_b\delta_{\delta r,0}+M_a\delta_{\delta r,-1} &    0 \\
    \end{pmatrix}\equiv(h'_{\delta r})_{s_1 s_2}\,.
\end{equation}
Exploiting the translational invariance in time and space coordinates, we define two types of Fourier transform for the Green's function
\begin{equation}
\label{eqSM:FTperiodicchain}
   G^{s_1s_2}_{r+\delta r, r}(\tau)=\frac{2}{L}\sum_{k=-L/4}^{L/4-1} G^{s_1s_2}_{k}(\tau)\,e^{-2\pi\mathrm{i}\frac{k\delta r}{L/2}} \,,
    \qquad
    G^{s_1s_2}_{k}(\tau)=\frac{1}{2\pi}\int_{-\infty}^\infty  G_k(\omega)e^{-\mathrm{i}\omega\tau} d\omega
    \,,
\end{equation}
where we have exploited the fact that, due to the translational invariance, the entries of the Green's function depend only on $r_1-r_2=\delta r$ and therefore the additional label $r$ is unimportant. Analogous transforms are defined for the self-energy $\Sigma_{xy}(\tau)=2\mathrm{i}h'_{xy}(G_{xy}(0))\delta(\tau)$. In this latter case, using (\ref{eqSM:potentialMatrix}), we have
\begin{equation}
\label{eqSM:self-energyperiodic}
    \Sigma^{s_1s_2}_{k}(\omega)= 2\mathrm{i}\begin{pmatrix}
        0   &   M_b-M_ae^{\mathrm{i}4\pi k/L} \\
        -M_b+M_a e^{\mathrm{i} 4\pi k/L} &    0 \\
    \end{pmatrix}
    =
    \mathrm{i}\begin{pmatrix}
        0   &   \varepsilon_k e^{\mathrm{i}\varphi_k} \\
       -\varepsilon_k e^{-\mathrm{i}\varphi_k} &    0 \\
    \end{pmatrix}\,,
\end{equation}
where we have introduced the parametrization
\begin{equation}
   \label{eqSM:radialparametrization}\varepsilon_k\equiv \sqrt{M_b^2+M_a^2-2M_aM_b\cos(4\pi k/L)} \,,
    \qquad
    e^{\mathrm{i}\varphi_k}\equiv
    \frac{M_b -M_a e^{\mathrm{i} 4\pi k/L}}{M_b-M_a e^{-\mathrm{i}4\pi k/L}}
    \,.
\end{equation}
The SD equations (3) of the main text in Fourier space (along both the space and time direction) read
\begin{equation}
\label{eqSM:SDFourierspacetime}
    G_k(\omega)=\frac{\mathrm{i}}{\omega+2h'_{\delta r}}=\mathrm{i}\frac{\omega+2h'_{\delta r}}{\omega^2+\varepsilon_k^2}\,,
\end{equation}
where $\omega$ is to be thought as being multiplied by the $4\times 4$ identity matrix, and in the last step we have used (\ref{eqSM:potentialMatrix}) and the parametrization in (\ref{eqSM:self-energyperiodic}). Assuming $\tau>0$ and inserting (\ref{eqSM:SDFourierspacetime}) into the second equation of (\ref{eqSM:FTperiodicchain}), we obtain
\begin{equation}
     G^{s_1s_2}_k(\tau)=\frac{e^{-\tau \varepsilon_k}}{2}
     \begin{pmatrix}
        1   &   \mathrm{i} e^{\mathrm{i}\varphi_k} \\
       -\mathrm{i} e^{-\mathrm{i}\varphi_k} &    1 \\
    \end{pmatrix}\,.
\end{equation}
Exploiting also the first equation of (\ref{eqSM:FTperiodicchain}), the Green's functions of the system in its ground state for $\tau>0$ read
\begin{equation}
\label{eqSM:two-point-periodic}
 G^{s_1s_2}_{r+\delta r, r}(\tau) =\frac{1}{L}\sum_{k=-L/4}^{L/4-1}e^{-\tau \varepsilon_k-4\pi\mathrm{i}\frac{k\delta r}{L}}\begin{pmatrix}
        1   &   \mathrm{i} e^{\mathrm{i}\varphi_k} \\
       -\mathrm{i} e^{-\mathrm{i}\varphi_k} &    1 \\
    \end{pmatrix}\longrightarrow\int_{-\frac{1}{2}}^\frac{1}{2} d p \frac{e^{-\tau \varepsilon(p)-2\pi\mathrm{i} p\delta r}}{2} \begin{pmatrix}
        1   &   \mathrm{i} e^{\mathrm{i}\varphi(p)} \\
       -\mathrm{i} e^{-\mathrm{i}\varphi(p)} &    1 \\
    \end{pmatrix} \,,
\end{equation}
where the last expression is obtained in the limit $L\to\infty$ of infinitely many sites along the chain integrating over $p=2k/L$, and we have introduced $\varepsilon(p)\equiv\lim_{L\to\infty} \varepsilon_{k=p L/2} $ and $\varphi(p)\equiv\lim_{L\to\infty} \varphi_{k=p L/2}$. The functions $\varepsilon(p)$ and $\varphi(p)$ in (\ref{eqSM:two-point-periodic}) are not explicitly known given that they depend via $M_a$ and $M_b$ on some of the entries of the Green's function evaluated at $\tau=0$. Thus, in order to determine the two-point functions and solve the model, we still need to find expressions of $\varepsilon(p)$ and $\varphi(p)$ in terms of the parameters of the microscopic Hamiltonian. Following the spirit of this work, we achieve this goal through a SC condition. We first notice that, from \eqref{eqSM:Ma_Mb_periodic},  $M_a$ and $M_b$ each only depend on a single but different component of the Green's function at $\tau=0$. Thus, we can focus on their form, which, from (\ref{eqSM:two-point-periodic}), is given by
\begin{equation}
\label{eqSM:GevenGodd}
   G_{r,r}^{0,1}(0)=\frac{\mathrm{i}}{2}\int_{-\frac{1}{2}}^\frac{1}{2} d p e^{\mathrm{i}\varphi(p)}=G_{2x,2x+1}(0)\,,
   \qquad
   G_{r,r+1}^{1,0}(0)=-\frac{\mathrm{i}}{2}\int_{-\frac{1}{2}}^\frac{1}{2} d p e^{2\pi\mathrm{i} p-\mathrm{i}\varphi(p)}=G_{2x-1,2x}(0)\,,
\end{equation}
where in the last steps we have restored the dependence of spatial coordinates on a single parameter. By changing integration variable $p\to p+1/2$ and exploiting (\ref{eqSM:radialparametrization}), we have
\begin{equation}
\label{eqSM:GevenGodd_v2}
    -2 \mathrm{i} G_{2x,2x+1}(0)=\int_0^1d p \frac{1-R\cos \kc{2 \pi  p }}{\sqrt{R^2-2 R \cos (2 \pi  p )+1}}\,,\qquad -2 \mathrm{i}G_{2x-1,2x}(0)=\int_0^1 d p \frac{R-\cos (2 \pi  p )}{\sqrt{R^2-2 R \cos (2 \pi  p )+1}}\,,
\end{equation}
where
\begin{equation}
R\equiv\frac{M_a}{M_b}=\frac{h'_a\left(-2 \mathrm{i}G_{2x-1,2x}(0)\right)}{h'_b\left(-2 \mathrm{i} G_{2x,2x+1}(0)\right)}\,.    
\end{equation}
By introducing $v$ such that $R=\frac{1-v}{1+v}$, (\ref{eqSM:GevenGodd_v2}) gives Eqs.~(15) and (16) of the main text. The problem is solved by inserting (15) into (14) of the main text, which yields an equation for $v$. The solution to the latter determines $G_{2x,2x+1}(0)$ and $G_{2x-1,2x}(0)$, which, in turn, are used to obtained the functions $\varepsilon(p)$ and $\varphi(p)$. Inserting $\varepsilon(p)$ and $\varphi(p)$ into (\ref{eqSM:two-point-periodic}), we can finally compute all the two-point functions, which, due to the large $N$ factorization, amounts to solving the model.

\bibliography{referencesPRL}

\begin{thebibliography}{44}%
\makeatletter
\providecommand \@ifxundefined [1]{%
 \@ifx{#1\undefined}
}%
\providecommand \@ifnum [1]{%
 \ifnum #1\expandafter \@firstoftwo
 \else \expandafter \@secondoftwo
 \fi
}%
\providecommand \@ifx [1]{%
 \ifx #1\expandafter \@firstoftwo
 \else \expandafter \@secondoftwo
 \fi
}%
\providecommand \natexlab [1]{#1}%
\providecommand \enquote  [1]{``#1''}%
\providecommand \bibnamefont  [1]{#1}%
\providecommand \bibfnamefont [1]{#1}%
\providecommand \citenamefont [1]{#1}%
\providecommand \href@noop [0]{\@secondoftwo}%
\providecommand \href [0]{\begingroup \@sanitize@url \@href}%
\providecommand \@href[1]{\@@startlink{#1}\@@href}%
\providecommand \@@href[1]{\endgroup#1\@@endlink}%
\providecommand \@sanitize@url [0]{\catcode `\\12\catcode `\$12\catcode `\&12\catcode `\#12\catcode `\^12\catcode `\_12\catcode `\%12\relax}%
\providecommand \@@startlink[1]{}%
\providecommand \@@endlink[0]{}%
\providecommand \url  [0]{\begingroup\@sanitize@url \@url }%
\providecommand \@url [1]{\endgroup\@href {#1}{\urlprefix }}%
\providecommand \urlprefix  [0]{URL }%
\providecommand \Eprint [0]{\href }%
\providecommand \doibase [0]{https://doi.org/}%
\providecommand \selectlanguage [0]{\@gobble}%
\providecommand \bibinfo  [0]{\@secondoftwo}%
\providecommand \bibfield  [0]{\@secondoftwo}%
\providecommand \translation [1]{[#1]}%
\providecommand \BibitemOpen [0]{}%
\providecommand \bibitemStop [0]{}%
\providecommand \bibitemNoStop [0]{.\EOS\space}%
\providecommand \EOS [0]{\spacefactor3000\relax}%
\providecommand \BibitemShut  [1]{\csname bibitem#1\endcsname}%
\let\auto@bib@innerbib\@empty
\bibitem [{\citenamefont {Murray}\ and\ \citenamefont {von Neumann}(1936)}]{Murray1936}%
  \BibitemOpen
  \bibfield  {author} {\bibinfo {author} {\bibfnamefont {F.~J.}\ \bibnamefont {Murray}}\ and\ \bibinfo {author} {\bibfnamefont {J.}~\bibnamefont {von Neumann}},\ }\bibfield  {title} {\bibinfo {title} {{On Rings of Operators}},\ }\href {https://doi.org/10.2307/1968693} {\bibfield  {journal} {\bibinfo  {journal} {The Annals of Mathematics}\ }\textbf {\bibinfo {volume} {37}},\ \bibinfo {pages} {116} (\bibinfo {year} {1936})}\BibitemShut {NoStop}%
\bibitem [{\citenamefont {Murray}\ and\ \citenamefont {von Neumann}(1937)}]{Murray1937}%
  \BibitemOpen
  \bibfield  {author} {\bibinfo {author} {\bibfnamefont {F.~J.}\ \bibnamefont {Murray}}\ and\ \bibinfo {author} {\bibfnamefont {J.}~\bibnamefont {von Neumann}},\ }\bibfield  {title} {\bibinfo {title} {{On rings of operators. II}},\ }\href {https://doi.org/10.1090/s0002-9947-1937-1501899-4} {\bibfield  {journal} {\bibinfo  {journal} {Transactions of the American Mathematical Society}\ }\textbf {\bibinfo {volume} {41}},\ \bibinfo {pages} {208–248} (\bibinfo {year} {1937})}\BibitemShut {NoStop}%
\bibitem [{\citenamefont {von Neumann}(1940)}]{vNeumann1940}%
  \BibitemOpen
  \bibfield  {author} {\bibinfo {author} {\bibfnamefont {J.}~\bibnamefont {von Neumann}},\ }\bibfield  {title} {\bibinfo {title} {{On Rings of Operators. III}},\ }\href {https://doi.org/10.2307/1968823} {\bibfield  {journal} {\bibinfo  {journal} {The Annals of Mathematics}\ }\textbf {\bibinfo {volume} {41}},\ \bibinfo {pages} {94} (\bibinfo {year} {1940})}\BibitemShut {NoStop}%
\bibitem [{\citenamefont {Murray}\ and\ \citenamefont {von Neumann}(1943)}]{Murray1943}%
  \BibitemOpen
  \bibfield  {author} {\bibinfo {author} {\bibfnamefont {F.~J.}\ \bibnamefont {Murray}}\ and\ \bibinfo {author} {\bibfnamefont {J.}~\bibnamefont {von Neumann}},\ }\bibfield  {title} {\bibinfo {title} {{On Rings of Operators. IV}},\ }\href {https://doi.org/10.2307/1969107} {\bibfield  {journal} {\bibinfo  {journal} {The Annals of Mathematics}\ }\textbf {\bibinfo {volume} {44}},\ \bibinfo {pages} {716} (\bibinfo {year} {1943})}\BibitemShut {NoStop}%
\bibitem [{\citenamefont {Takesaki}(1979)}]{Takesaki1979}%
  \BibitemOpen
  \bibfield  {author} {\bibinfo {author} {\bibfnamefont {M.}~\bibnamefont {Takesaki}},\ }\href {https://doi.org/10.1007/978-1-4612-6188-9} {\emph {\bibinfo {title} {{Theory of Operator Algebras I-III}}}}\ (\bibinfo  {publisher} {Springer New York},\ \bibinfo {year} {1979})\BibitemShut {NoStop}%
\bibitem [{\citenamefont {Haag}(1996)}]{Haag1996}%
  \BibitemOpen
  \bibfield  {author} {\bibinfo {author} {\bibfnamefont {R.}~\bibnamefont {Haag}},\ }\href {https://doi.org/10.1007/978-3-642-61458-3} {\emph {\bibinfo {title} {{Local Quantum Physics}}}}\ (\bibinfo  {publisher} {Springer Berlin Heidelberg},\ \bibinfo {year} {1996})\BibitemShut {NoStop}%
\bibitem [{\citenamefont {Witten}(2018)}]{Witten:2018zxz}%
  \BibitemOpen
  \bibfield  {author} {\bibinfo {author} {\bibfnamefont {E.}~\bibnamefont {Witten}},\ }\bibfield  {title} {\bibinfo {title} {{APS Medal for Exceptional Achievement in Research: Invited article on entanglement properties of quantum field theory}},\ }\href {https://doi.org/10.1103/RevModPhys.90.045003} {\bibfield  {journal} {\bibinfo  {journal} {Rev. Mod. Phys.}\ }\textbf {\bibinfo {volume} {90}},\ \bibinfo {pages} {045003} (\bibinfo {year} {2018})},\ \Eprint {https://arxiv.org/abs/1803.04993} {arXiv:1803.04993 [hep-th]} \BibitemShut {NoStop}%
\bibitem [{\citenamefont {Sorce}(2023)}]{Sorce:2023fdx}%
  \BibitemOpen
  \bibfield  {author} {\bibinfo {author} {\bibfnamefont {J.}~\bibnamefont {Sorce}},\ }\href@noop {} {\bibinfo {title} {{Notes on the type classification of von Neumann algebras}}} (\bibinfo {year} {2023}),\ \Eprint {https://arxiv.org/abs/2302.01958} {arXiv:2302.01958 [hep-th]} \BibitemShut {NoStop}%
\bibitem [{\citenamefont {Maldacena}(1999)}]{Maldacena1997}%
  \BibitemOpen
  \bibfield  {author} {\bibinfo {author} {\bibfnamefont {J.}~\bibnamefont {Maldacena}},\ }\bibfield  {title} {\bibinfo {title} {{The Large-N Limit of Superconformal Field Theories and Supergravity}},\ }\href {https://doi.org/10.1023/A:1026654312961} {\bibfield  {journal} {\bibinfo  {journal} {International Journal of Theoretical Physics}\ }\textbf {\bibinfo {volume} {38}},\ \bibinfo {pages} {1113} (\bibinfo {year} {1999})},\ \Eprint {https://arxiv.org/abs/hep-th/9711200} {arXiv:hep-th/9711200 [hep-th]} \BibitemShut {NoStop}%
\bibitem [{\citenamefont {Witten}(1998)}]{Witten}%
  \BibitemOpen
  \bibfield  {author} {\bibinfo {author} {\bibfnamefont {E.}~\bibnamefont {Witten}},\ }\bibfield  {title} {\bibinfo {title} {{Anti-de Sitter space and holography}},\ }\href {https://doi.org/10.4310/ATMP.1998.v2.n2.a2} {\bibfield  {journal} {\bibinfo  {journal} {Adv. Theor. Math. Phys.}\ }\textbf {\bibinfo {volume} {2}},\ \bibinfo {pages} {253} (\bibinfo {year} {1998})},\ \Eprint {https://arxiv.org/abs/hep-th/9802150} {arXiv:hep-th/9802150} \BibitemShut {NoStop}%
\bibitem [{\citenamefont {Gubser}\ \emph {et~al.}(1998)\citenamefont {Gubser}, \citenamefont {Klebanov},\ and\ \citenamefont {Polyakov}}]{GubserKlebanovPolyakov}%
  \BibitemOpen
  \bibfield  {author} {\bibinfo {author} {\bibfnamefont {S.~S.}\ \bibnamefont {Gubser}}, \bibinfo {author} {\bibfnamefont {I.~R.}\ \bibnamefont {Klebanov}},\ and\ \bibinfo {author} {\bibfnamefont {A.~M.}\ \bibnamefont {Polyakov}},\ }\bibfield  {title} {\bibinfo {title} {{Gauge theory correlators from noncritical string theory}},\ }\href {https://doi.org/10.1016/S0370-2693(98)00377-3} {\bibfield  {journal} {\bibinfo  {journal} {Phys. Lett.}\ }\textbf {\bibinfo {volume} {B428}},\ \bibinfo {pages} {105} (\bibinfo {year} {1998})},\ \Eprint {https://arxiv.org/abs/hep-th/9802109} {arXiv:hep-th/9802109} \BibitemShut {NoStop}%
\bibitem [{\citenamefont {Leutheusser}\ and\ \citenamefont {Liu}(2023{\natexlab{a}})}]{Leutheusser:2021qhd}%
  \BibitemOpen
  \bibfield  {author} {\bibinfo {author} {\bibfnamefont {S.}~\bibnamefont {Leutheusser}}\ and\ \bibinfo {author} {\bibfnamefont {H.}~\bibnamefont {Liu}},\ }\bibfield  {title} {\bibinfo {title} {{Causal connectability between quantum systems and the black hole interior in holographic duality}},\ }\href {https://doi.org/10.1103/PhysRevD.108.086019} {\bibfield  {journal} {\bibinfo  {journal} {Phys. Rev. D}\ }\textbf {\bibinfo {volume} {108}},\ \bibinfo {pages} {086019} (\bibinfo {year} {2023}{\natexlab{a}})},\ \Eprint {https://arxiv.org/abs/2110.05497} {arXiv:2110.05497 [hep-th]} \BibitemShut {NoStop}%
\bibitem [{\citenamefont {Leutheusser}\ and\ \citenamefont {Liu}(2023{\natexlab{b}})}]{Leutheusser:2021frk}%
  \BibitemOpen
  \bibfield  {author} {\bibinfo {author} {\bibfnamefont {S.}~\bibnamefont {Leutheusser}}\ and\ \bibinfo {author} {\bibfnamefont {H.}~\bibnamefont {Liu}},\ }\bibfield  {title} {\bibinfo {title} {{Emergent times in holographic duality}},\ }\href {https://doi.org/10.1103/PhysRevD.108.086020} {\bibfield  {journal} {\bibinfo  {journal} {Phys. Rev. D}\ }\textbf {\bibinfo {volume} {108}},\ \bibinfo {pages} {086020} (\bibinfo {year} {2023}{\natexlab{b}})},\ \Eprint {https://arxiv.org/abs/2112.12156} {arXiv:2112.12156 [hep-th]} \BibitemShut {NoStop}%
\bibitem [{\citenamefont {Witten}(2022{\natexlab{a}})}]{Witten:2021jzq}%
  \BibitemOpen
  \bibfield  {author} {\bibinfo {author} {\bibfnamefont {E.}~\bibnamefont {Witten}},\ }\bibinfo {title} {{Why Does Quantum Field Theory in Curved Spacetime Make Sense? And What Happens to the Algebra of Observables in the Thermodynamic Limit?}},\ in\ \href {https://doi.org/10.1007/978-3-031-17523-7_11} {\emph {\bibinfo {booktitle} {Dialogues Between Physics and Mathematics}}}\ (\bibinfo  {publisher} {Springer International Publishing},\ \bibinfo {year} {2022})\ p.\ \bibinfo {pages} {241–284}\BibitemShut {NoStop}%
\bibitem [{\citenamefont {Witten}(2022{\natexlab{b}})}]{Witten:2021unn}%
  \BibitemOpen
  \bibfield  {author} {\bibinfo {author} {\bibfnamefont {E.}~\bibnamefont {Witten}},\ }\bibfield  {title} {\bibinfo {title} {{Gravity and the crossed product}},\ }\bibfield  {journal} {\bibinfo  {journal} {Journal of High Energy Physics}\ }\textbf {\bibinfo {volume} {2022}},\ \href {https://doi.org/10.1007/jhep10(2022)008} {10.1007/jhep10(2022)008} (\bibinfo {year} {2022}{\natexlab{b}})\BibitemShut {NoStop}%
\bibitem [{\citenamefont {Chandrasekaran}\ \emph {et~al.}(2023{\natexlab{a}})\citenamefont {Chandrasekaran}, \citenamefont {Penington},\ and\ \citenamefont {Witten}}]{Chandrasekaran:2022eqq}%
  \BibitemOpen
  \bibfield  {author} {\bibinfo {author} {\bibfnamefont {V.}~\bibnamefont {Chandrasekaran}}, \bibinfo {author} {\bibfnamefont {G.}~\bibnamefont {Penington}},\ and\ \bibinfo {author} {\bibfnamefont {E.}~\bibnamefont {Witten}},\ }\bibfield  {title} {\bibinfo {title} {{Large \text{N} algebras and generalized entropy}},\ }\bibfield  {journal} {\bibinfo  {journal} {Journal of High Energy Physics}\ }\textbf {\bibinfo {volume} {2023}},\ \href {https://doi.org/10.1007/jhep04(2023)009} {10.1007/jhep04(2023)009} (\bibinfo {year} {2023}{\natexlab{a}})\BibitemShut {NoStop}%
\bibitem [{\citenamefont {Chandrasekaran}\ \emph {et~al.}(2023{\natexlab{b}})\citenamefont {Chandrasekaran}, \citenamefont {Longo}, \citenamefont {Penington},\ and\ \citenamefont {Witten}}]{Chandrasekaran:2022cip}%
  \BibitemOpen
  \bibfield  {author} {\bibinfo {author} {\bibfnamefont {V.}~\bibnamefont {Chandrasekaran}}, \bibinfo {author} {\bibfnamefont {R.}~\bibnamefont {Longo}}, \bibinfo {author} {\bibfnamefont {G.}~\bibnamefont {Penington}},\ and\ \bibinfo {author} {\bibfnamefont {E.}~\bibnamefont {Witten}},\ }\bibfield  {title} {\bibinfo {title} {{An algebra of observables for de Sitter space}},\ }\bibfield  {journal} {\bibinfo  {journal} {Journal of High Energy Physics}\ }\textbf {\bibinfo {volume} {2023}},\ \href {https://doi.org/10.1007/jhep02(2023)082} {10.1007/jhep02(2023)082} (\bibinfo {year} {2023}{\natexlab{b}})\BibitemShut {NoStop}%
\bibitem [{\citenamefont {Banerjee}\ \emph {et~al.}(2023)\citenamefont {Banerjee}, \citenamefont {Dorband}, \citenamefont {Erdmenger},\ and\ \citenamefont {Weigel}}]{Banerjee:2023eew}%
  \BibitemOpen
  \bibfield  {author} {\bibinfo {author} {\bibfnamefont {S.}~\bibnamefont {Banerjee}}, \bibinfo {author} {\bibfnamefont {M.}~\bibnamefont {Dorband}}, \bibinfo {author} {\bibfnamefont {J.}~\bibnamefont {Erdmenger}},\ and\ \bibinfo {author} {\bibfnamefont {A.-L.}\ \bibnamefont {Weigel}},\ }\bibfield  {title} {\bibinfo {title} {{Geometric phases characterise operator algebras and missing information}},\ }\href {https://doi.org/10.1007/JHEP10(2023)026} {\bibfield  {journal} {\bibinfo  {journal} {JHEP}\ }\textbf {\bibinfo {volume} {10}},\ \bibinfo {pages} {026}},\ \Eprint {https://arxiv.org/abs/2306.00055} {arXiv:2306.00055 [hep-th]} \BibitemShut {NoStop}%
\bibitem [{\citenamefont {Engelhardt}\ and\ \citenamefont {Liu}(2023)}]{Engelhardt:2023xer}%
  \BibitemOpen
  \bibfield  {author} {\bibinfo {author} {\bibfnamefont {N.}~\bibnamefont {Engelhardt}}\ and\ \bibinfo {author} {\bibfnamefont {H.}~\bibnamefont {Liu}},\ }\bibfield  {title} {\bibinfo {title} {{Algebraic ER=EPR and Complexity Transfer}},\ }\href@noop {} {\  (\bibinfo {year} {2023})},\ \Eprint {https://arxiv.org/abs/2311.04281} {arXiv:2311.04281 [hep-th]} \BibitemShut {NoStop}%
\bibitem [{\citenamefont {Peschel}(2003)}]{IngoPeschel_2003}%
  \BibitemOpen
  \bibfield  {author} {\bibinfo {author} {\bibfnamefont {I.}~\bibnamefont {Peschel}},\ }\bibfield  {title} {\bibinfo {title} {{Calculation of reduced density matrices from correlation functions}},\ }\href {https://doi.org/10.1088/0305-4470/36/14/10} {\bibfield  {journal} {\bibinfo  {journal} {J. Phys. A}\ }\textbf {\bibinfo {volume} {36}},\ \bibinfo {pages} {L205} (\bibinfo {year} {2003})}\BibitemShut {NoStop}%
\bibitem [{\citenamefont {Basteiro}\ \emph {et~al.}(2023)\citenamefont {Basteiro}, \citenamefont {Das}, \citenamefont {Giulio},\ and\ \citenamefont {Erdmenger}}]{Basteiro2023}%
  \BibitemOpen
  \bibfield  {author} {\bibinfo {author} {\bibfnamefont {P.}~\bibnamefont {Basteiro}}, \bibinfo {author} {\bibfnamefont {R.~N.}\ \bibnamefont {Das}}, \bibinfo {author} {\bibfnamefont {G.~D.}\ \bibnamefont {Giulio}},\ and\ \bibinfo {author} {\bibfnamefont {J.}~\bibnamefont {Erdmenger}},\ }\bibfield  {title} {\bibinfo {title} {{Aperiodic spin chains at the boundary of hyperbolic tilings}},\ }\href {https://doi.org/10.21468/SciPostPhys.15.5.218} {\bibfield  {journal} {\bibinfo  {journal} {SciPost Phys.}\ }\textbf {\bibinfo {volume} {15}},\ \bibinfo {pages} {218} (\bibinfo {year} {2023})}\BibitemShut {NoStop}%
\bibitem [{\citenamefont {Gross}\ and\ \citenamefont {Neveu}(1974)}]{PhysRevD.10.3235}%
  \BibitemOpen
  \bibfield  {author} {\bibinfo {author} {\bibfnamefont {D.~J.}\ \bibnamefont {Gross}}\ and\ \bibinfo {author} {\bibfnamefont {A.}~\bibnamefont {Neveu}},\ }\bibfield  {title} {\bibinfo {title} {{Dynamical symmetry breaking in asymptotically free field theories}},\ }\href {https://doi.org/10.1103/PhysRevD.10.3235} {\bibfield  {journal} {\bibinfo  {journal} {Phys. Rev. D}\ }\textbf {\bibinfo {volume} {10}},\ \bibinfo {pages} {3235} (\bibinfo {year} {1974})}\BibitemShut {NoStop}%
\bibitem [{Note1()}]{Note1}%
  \BibitemOpen
  \bibinfo {note} {See Supplemental Material [url] for more details.}\BibitemShut {Stop}%
\bibitem [{\citenamefont {Saad}\ \emph {et~al.}(2018)\citenamefont {Saad}, \citenamefont {Shenker},\ and\ \citenamefont {Stanford}}]{SSS18rampSYK}%
  \BibitemOpen
  \bibfield  {author} {\bibinfo {author} {\bibfnamefont {P.}~\bibnamefont {Saad}}, \bibinfo {author} {\bibfnamefont {S.~H.}\ \bibnamefont {Shenker}},\ and\ \bibinfo {author} {\bibfnamefont {D.}~\bibnamefont {Stanford}},\ }\bibfield  {title} {\bibinfo {title} {{A semiclassical ramp in SYK and in gravity}},\ }\href@noop {} {\bibfield  {journal} {\bibinfo  {journal} {\hspace{0pt}}\ } (\bibinfo {year} {2018})},\ \Eprint {https://arxiv.org/abs/1806.06840} {arXiv:1806.06840 [hep-th]} \BibitemShut {NoStop}%
\bibitem [{\citenamefont {Sachdev}\ and\ \citenamefont {Ye}(1993)}]{SachdevYePRL1993}%
  \BibitemOpen
  \bibfield  {author} {\bibinfo {author} {\bibfnamefont {S.}~\bibnamefont {Sachdev}}\ and\ \bibinfo {author} {\bibfnamefont {J.}~\bibnamefont {Ye}},\ }\bibfield  {title} {\bibinfo {title} {{Gapless spin-fluid ground state in a random quantum Heisenberg magnet}},\ }\href {https://doi.org/10.1103/PhysRevLett.70.3339} {\bibfield  {journal} {\bibinfo  {journal} {Phys. Rev. Lett.}\ }\textbf {\bibinfo {volume} {70}},\ \bibinfo {pages} {3339} (\bibinfo {year} {1993})}\BibitemShut {NoStop}%
\bibitem [{\citenamefont {Kitaev}(2015)}]{kitaev}%
  \BibitemOpen
  \bibfield  {author} {\bibinfo {author} {\bibfnamefont {A.}~\bibnamefont {Kitaev}},\ }\href {https://online.kitp.ucsb.edu/online/entangled15/kitaev/} {\bibinfo {title} {{A simple model of quantum holography}}} (\bibinfo {year} {2015}),\ \bibinfo {note} {talks at KITP, April 7 and May 27.}\BibitemShut {Stop}%
\bibitem [{\citenamefont {Maldacena}\ and\ \citenamefont {Stanford}(2016)}]{Maldacena:2016hyu}%
  \BibitemOpen
  \bibfield  {author} {\bibinfo {author} {\bibfnamefont {J.}~\bibnamefont {Maldacena}}\ and\ \bibinfo {author} {\bibfnamefont {D.}~\bibnamefont {Stanford}},\ }\bibfield  {title} {\bibinfo {title} {{Remarks on the Sachdev-Ye-Kitaev model}},\ }\href {https://doi.org/10.1103/PhysRevD.94.106002} {\bibfield  {journal} {\bibinfo  {journal} {Phys. Rev. D}\ }\textbf {\bibinfo {volume} {94}},\ \bibinfo {pages} {106002} (\bibinfo {year} {2016})},\ \Eprint {https://arxiv.org/abs/1604.07818} {arXiv:1604.07818 [hep-th]} \BibitemShut {NoStop}%
\bibitem [{\citenamefont {Sachdev}(1999)}]{Sachdev_1999}%
  \BibitemOpen
  \bibfield  {author} {\bibinfo {author} {\bibfnamefont {S.}~\bibnamefont {Sachdev}},\ }\bibfield  {title} {\bibinfo {title} {{Quantum phase transitions}},\ }\href {https://doi.org/10.1088/2058-7058/12/4/23} {\bibfield  {journal} {\bibinfo  {journal} {Physics World}\ }\textbf {\bibinfo {volume} {12}},\ \bibinfo {pages} {33} (\bibinfo {year} {1999})}\BibitemShut {NoStop}%
\bibitem [{\citenamefont {von Neumann}(1939)}]{vNeumann1939}%
  \BibitemOpen
  \bibfield  {author} {\bibinfo {author} {\bibfnamefont {J.}~\bibnamefont {von Neumann}},\ }\bibfield  {title} {\bibinfo {title} {{On infinite direct products}},\ }\href {http://www.numdam.org/item/CM_1939__6__1_0.pdf} {\bibfield  {journal} {\bibinfo  {journal} {Compositio Mathematica}\ }\textbf {\bibinfo {volume} {6}},\ \bibinfo {pages} {1} (\bibinfo {year} {1939})}\BibitemShut {NoStop}%
\bibitem [{\citenamefont {Powers}(1967)}]{Powers1967}%
  \BibitemOpen
  \bibfield  {author} {\bibinfo {author} {\bibfnamefont {R.~T.}\ \bibnamefont {Powers}},\ }\bibfield  {title} {\bibinfo {title} {{Representations of Uniformly Hyperfinite Algebras and Their Associated von Neumann Rings}},\ }\href {https://doi.org/10.2307/1970364} {\bibfield  {journal} {\bibinfo  {journal} {The Annals of Mathematics}\ }\textbf {\bibinfo {volume} {86}},\ \bibinfo {pages} {138} (\bibinfo {year} {1967})}\BibitemShut {NoStop}%
\bibitem [{\citenamefont {Araki}\ and\ \citenamefont {Woods}(1968)}]{Araki1968}%
  \BibitemOpen
  \bibfield  {author} {\bibinfo {author} {\bibfnamefont {H.}~\bibnamefont {Araki}}\ and\ \bibinfo {author} {\bibfnamefont {E.~J.}\ \bibnamefont {Woods}},\ }\bibfield  {title} {\bibinfo {title} {{A classification of factors}},\ }\href {https://doi.org/10.2977/prims/1195195263} {\bibfield  {journal} {\bibinfo  {journal} {Publications of the Research Institute for Mathematical Sciences}\ }\textbf {\bibinfo {volume} {4}},\ \bibinfo {pages} {51–130} (\bibinfo {year} {1968})}\BibitemShut {NoStop}%
\bibitem [{\citenamefont {Pedersen}(2018)}]{Pedersen2018}%
  \BibitemOpen
  \bibfield  {author} {\bibinfo {author} {\bibfnamefont {G.~K.}\ \bibnamefont {Pedersen}},\ }\href {https://doi.org/10.1016/b978-0-12-814122-9.00007-6} {\emph {\bibinfo {title} {$C^*$-Algebras and their Automorphism Groups}}}\ (\bibinfo  {publisher} {Elsevier},\ \bibinfo {year} {2018})\BibitemShut {NoStop}%
\bibitem [{\citenamefont {Sachdev}(2015)}]{Sachdev:2015efa}%
  \BibitemOpen
  \bibfield  {author} {\bibinfo {author} {\bibfnamefont {S.}~\bibnamefont {Sachdev}},\ }\bibfield  {title} {\bibinfo {title} {{Bekenstein-Hawking Entropy and Strange Metals}},\ }\href {https://doi.org/10.1103/PhysRevX.5.041025} {\bibfield  {journal} {\bibinfo  {journal} {Phys. Rev. X}\ }\textbf {\bibinfo {volume} {5}},\ \bibinfo {pages} {041025} (\bibinfo {year} {2015})},\ \Eprint {https://arxiv.org/abs/1506.05111} {arXiv:1506.05111 [hep-th]} \BibitemShut {NoStop}%
\bibitem [{\citenamefont {Gu}\ \emph {et~al.}(2017)\citenamefont {Gu}, \citenamefont {Qi},\ and\ \citenamefont {Stanford}}]{Gu:2016oyy}%
  \BibitemOpen
  \bibfield  {author} {\bibinfo {author} {\bibfnamefont {Y.}~\bibnamefont {Gu}}, \bibinfo {author} {\bibfnamefont {X.-L.}\ \bibnamefont {Qi}},\ and\ \bibinfo {author} {\bibfnamefont {D.}~\bibnamefont {Stanford}},\ }\bibfield  {title} {\bibinfo {title} {{Local criticality, diffusion and chaos in generalized Sachdev-Ye-Kitaev models}},\ }\href {https://doi.org/10.1007/JHEP05(2017)125} {\bibfield  {journal} {\bibinfo  {journal} {JHEP}\ }\bibfield  {number} {\bibinfo  {number} { (05)},\ \bibinfo {pages} {125}},\ }\Eprint {https://arxiv.org/abs/1609.07832} {arXiv:1609.07832 [hep-th]} \BibitemShut {NoStop}%
\bibitem [{\citenamefont {Maldacena}\ and\ \citenamefont {Qi}(2018)}]{Maldacena:2018lmt}%
  \BibitemOpen
  \bibfield  {author} {\bibinfo {author} {\bibfnamefont {J.}~\bibnamefont {Maldacena}}\ and\ \bibinfo {author} {\bibfnamefont {X.-L.}\ \bibnamefont {Qi}},\ }\href@noop {} {\bibinfo {title} {{Eternal traversable wormhole}}} (\bibinfo {year} {2018}),\ \Eprint {https://arxiv.org/abs/1804.00491} {arXiv:1804.00491 [hep-th]} \BibitemShut {NoStop}%
\bibitem [{\citenamefont {Numasawa}(2022)}]{Numasawa:2020sty}%
  \BibitemOpen
  \bibfield  {author} {\bibinfo {author} {\bibfnamefont {T.}~\bibnamefont {Numasawa}},\ }\bibfield  {title} {\bibinfo {title} {{Four coupled SYK models and nearly AdS$_{2}$ gravities: phase transitions in traversable wormholes and in bra-ket wormholes}},\ }\href {https://doi.org/10.1088/1361-6382/ac5736} {\bibfield  {journal} {\bibinfo  {journal} {Class. Quant. Grav.}\ }\textbf {\bibinfo {volume} {39}},\ \bibinfo {pages} {084001} (\bibinfo {year} {2022})},\ \Eprint {https://arxiv.org/abs/2011.12962} {arXiv:2011.12962 [hep-th]} \BibitemShut {NoStop}%
\bibitem [{\citenamefont {Stanford}\ \emph {et~al.}(2023)\citenamefont {Stanford}, \citenamefont {Vardhan},\ and\ \citenamefont {Yao}}]{StanfordScramblon23}%
  \BibitemOpen
  \bibfield  {author} {\bibinfo {author} {\bibfnamefont {D.}~\bibnamefont {Stanford}}, \bibinfo {author} {\bibfnamefont {S.}~\bibnamefont {Vardhan}},\ and\ \bibinfo {author} {\bibfnamefont {S.}~\bibnamefont {Yao}},\ }\bibfield  {title} {\bibinfo {title} {{Scramblon loops}},\ }\href@noop {} {\bibfield  {journal} {\bibinfo  {journal} {\,}\ } (\bibinfo {year} {2023})},\ \Eprint {https://arxiv.org/abs/2311.12121} {arXiv:2311.12121 [hep-th]} \BibitemShut {NoStop}%
\bibitem [{\citenamefont {Milekhin}\ and\ \citenamefont {Xu}(2023)}]{milekhin2023revisiting}%
  \BibitemOpen
  \bibfield  {author} {\bibinfo {author} {\bibfnamefont {A.}~\bibnamefont {Milekhin}}\ and\ \bibinfo {author} {\bibfnamefont {J.}~\bibnamefont {Xu}},\ }\href@noop {} {\bibinfo {title} {{Revisiting Brownian \text{SYK} and its possible relations to de Sitter}}} (\bibinfo {year} {2023}),\ \Eprint {https://arxiv.org/abs/2312.03623} {arXiv:2312.03623 [hep-th]} \BibitemShut {NoStop}%
\bibitem [{\citenamefont {Jian}\ and\ \citenamefont {Swingle}(2023)}]{Jian:2021tli}%
  \BibitemOpen
  \bibfield  {author} {\bibinfo {author} {\bibfnamefont {S.-K.}\ \bibnamefont {Jian}}\ and\ \bibinfo {author} {\bibfnamefont {B.}~\bibnamefont {Swingle}},\ }\bibfield  {title} {\bibinfo {title} {{Phase transition in von Neumann entropy from replica symmetry breaking}},\ }\href {https://doi.org/10.1007/JHEP11(2023)221} {\bibfield  {journal} {\bibinfo  {journal} {JHEP}\ }\textbf {\bibinfo {volume} {11}},\ \bibinfo {pages} {221}},\ \Eprint {https://arxiv.org/abs/2108.11973} {arXiv:2108.11973 [quant-ph]} \BibitemShut {NoStop}%
\bibitem [{\citenamefont {Vieira}(2005)}]{Vieira:2005PRB}%
  \BibitemOpen
  \bibfield  {author} {\bibinfo {author} {\bibfnamefont {A.~P.}\ \bibnamefont {Vieira}},\ }\bibfield  {title} {\bibinfo {title} {{Aperiodic quantum \textrm{XXZ} chains: Renormalization-group results}},\ }\href {https://doi.org/10.1103/PhysRevB.71.134408} {\bibfield  {journal} {\bibinfo  {journal} {Phys. Rev. B}\ }\textbf {\bibinfo {volume} {71}},\ \bibinfo {pages} {134408} (\bibinfo {year} {2005})}\BibitemShut {NoStop}%
\bibitem [{\citenamefont {Juh\'asz}\ and\ \citenamefont {Zimbor\'as}(2007)}]{Juh_sz_2007}%
  \BibitemOpen
  \bibfield  {author} {\bibinfo {author} {\bibfnamefont {R.}~\bibnamefont {Juh\'asz}}\ and\ \bibinfo {author} {\bibfnamefont {Z.}~\bibnamefont {Zimbor\'as}},\ }\bibfield  {title} {\bibinfo {title} {{Entanglement entropy in aperiodic singlet phases}},\ }\href {https://doi.org/10.1088/1742-5468/2007/04/p04004} {\bibfield  {journal} {\bibinfo  {journal} {J. Stat. Mech}\ }\textbf {\bibinfo {volume} {2007}},\ \bibinfo {pages} {P04004} (\bibinfo {year} {2007})},\ \Eprint {https://arxiv.org/abs/cond-mat/0703527} {arXiv:cond-mat/0703527 [cond-mat.stat-mech]} \BibitemShut {NoStop}%
\bibitem [{\citenamefont {Jahn}\ \emph {et~al.}(2020)\citenamefont {Jahn}, \citenamefont {Zimbor\'as},\ and\ \citenamefont {Eisert}}]{Jahn2020}%
  \BibitemOpen
  \bibfield  {author} {\bibinfo {author} {\bibfnamefont {A.}~\bibnamefont {Jahn}}, \bibinfo {author} {\bibfnamefont {Z.}~\bibnamefont {Zimbor\'as}},\ and\ \bibinfo {author} {\bibfnamefont {J.}~\bibnamefont {Eisert}},\ }\bibfield  {title} {\bibinfo {title} {{Central charges of aperiodic holographic tensor-network models}},\ }\href {https://doi.org/10.1103/PhysRevA.102.042407} {\bibfield  {journal} {\bibinfo  {journal} {Phys. Rev. A}\ }\textbf {\bibinfo {volume} {102}},\ \bibinfo {pages} {042407} (\bibinfo {year} {2020})}\BibitemShut {NoStop}%
\bibitem [{\citenamefont {Jahn}\ \emph {et~al.}(2022)\citenamefont {Jahn}, \citenamefont {Gluza}, \citenamefont {Verhoeven}, \citenamefont {Singh},\ and\ \citenamefont {Eisert}}]{Jahn:2021kti}%
  \BibitemOpen
  \bibfield  {author} {\bibinfo {author} {\bibfnamefont {A.}~\bibnamefont {Jahn}}, \bibinfo {author} {\bibfnamefont {M.}~\bibnamefont {Gluza}}, \bibinfo {author} {\bibfnamefont {C.}~\bibnamefont {Verhoeven}}, \bibinfo {author} {\bibfnamefont {S.}~\bibnamefont {Singh}},\ and\ \bibinfo {author} {\bibfnamefont {J.}~\bibnamefont {Eisert}},\ }\bibfield  {title} {\bibinfo {title} {{Boundary theories of critical matchgate tensor networks}},\ }\href {https://doi.org/10.1007/JHEP04(2022)111} {\bibfield  {journal} {\bibinfo  {journal} {JHEP}\ }\textbf {\bibinfo {volume} {04}},\ \bibinfo {pages} {111}},\ \Eprint {https://arxiv.org/abs/2110.02972} {arXiv:2110.02972 [quant-ph]} \BibitemShut {NoStop}%
\bibitem [{\citenamefont {Basteiro}\ \emph {et~al.}(2022)\citenamefont {Basteiro}, \citenamefont {Di~Giulio}, \citenamefont {Erdmenger}, \citenamefont {Karl}, \citenamefont {Meyer},\ and\ \citenamefont {Xian}}]{Basteiro:2022zur}%
  \BibitemOpen
  \bibfield  {author} {\bibinfo {author} {\bibfnamefont {P.}~\bibnamefont {Basteiro}}, \bibinfo {author} {\bibfnamefont {G.}~\bibnamefont {Di~Giulio}}, \bibinfo {author} {\bibfnamefont {J.}~\bibnamefont {Erdmenger}}, \bibinfo {author} {\bibfnamefont {J.}~\bibnamefont {Karl}}, \bibinfo {author} {\bibfnamefont {R.}~\bibnamefont {Meyer}},\ and\ \bibinfo {author} {\bibfnamefont {Z.-Y.}\ \bibnamefont {Xian}},\ }\bibfield  {title} {\bibinfo {title} {{Towards Explicit Discrete Holography: Aperiodic Spin Chains from Hyperbolic Tilings}},\ }\href {https://doi.org/10.21468/SciPostPhys.13.5.103} {\bibfield  {journal} {\bibinfo  {journal} {SciPost Phys.}\ }\textbf {\bibinfo {volume} {13}},\ \bibinfo {pages} {103} (\bibinfo {year} {2022})},\ \Eprint {https://arxiv.org/abs/2205.05693} {arXiv:2205.05693 [hep-th]} \BibitemShut {NoStop}%
\end{thebibliography}%

\end{document}